\documentclass[twocolumn,showpacs,preprintnumbers,nofootinbib,superscriptaddress]{revtex4-1}
\usepackage[utf8]{inputenc}
\usepackage[english]{babel}
\usepackage{microtype}	
\usepackage{xspace}
\usepackage{amsmath,amssymb,amsfonts,dsfont,bm} 	
\usepackage{hyperref}
\usepackage{graphicx}
\graphicspath{{figures/}{./}}
\usepackage[normalem]{ulem}
\usepackage{minibox}
\usepackage{color}
\usepackage{xcolor}

\newcommand{\be}{\begin{equation}}  
\newcommand{\ee}{\end{equation}}  
\newcommand{\bear}{\begin{eqnarray}}  
\newcommand{\eear}{\end{eqnarray}}  
\newcommand{\ba}{\begin{array}}  
	\newcommand{\ea}{\end{array}}  
\usepackage{bbold}

\newcommand{\abs}[1]{\left| #1 \right|}
\newcommand{\vev}[1]{\left< #1 \right>}

\usepackage{diagbox}

\newcommand{\no}{\nonumber}
\newcommand{\equaref}[1]{Eq.~(\ref{#1})}
\newcommand{\equasref}[2]{Eqs.~(\ref{#1})~and~(\ref{#2})}

\newcommand{\figref}[1]{Fig.~\ref{#1}}

\newcommand{\secref}[1]{Section~\ref{#1}}

\newcommand{\refref}[1]{Ref.~\cite{#1}}

\definecolor{rossoCP3}{cmyk}{0,.88,.77,.40}
\definecolor{blueRef}{rgb}{0.2,0.2,0.6}
\definecolor{blue}{rgb}{0,0.396,0.741}
\hypersetup{
	colorlinks, 
	bookmarksopen, 
	bookmarksnumbered,
	citecolor=blueRef, 		
	linkcolor=rossoCP3,	
	urlcolor=rossoCP3,			
}

\newskip\humongous \humongous=0pt plus 1000pt minus 1000pt

\newif\ifdtup

\def\oldreffmt#1{\rlap{[#1]} \hbox to 2\parindent{}}

\def\figfmt#1{\rlap{Figure {#1}} \hbox to 1in{}}  
  
\def\ie{\hbox{\it i.e.}{}\xspace }	\def\etc{\hbox{\it etc.}{}}  
\def\eg{\hbox{\it e.g.}{}\xspace }

\newcommand{\tabref}[1]{Table~\ref{#1}}

\def\vev#1{\left\langle #1\right\rangle}  
  
\def\abs#1{\left| #1\right|}

\def\beq{\begin{equation}}  
\def\eeq{\end{equation}}  
\def\bea{\begin{eqnarray}}  
\def\eea{\end{eqnarray}}

\def\bq{\begin{quote}}  
\def\eq{\end{quote}}

\def \lta {\mathrel{\vcenter  
     {\hbox{$<$}\nointerlineskip\hbox{$\sim$}}}}  
\def \gta {\mathrel{\vcenter  
     {\hbox{$>$}\nointerlineskip\hbox{$\sim$}}}}   

\relax  
\newcommand{\hc}{\; + \; \mathrm{h.c.} \;}
\newcommand{\LL}{\mathrm{L}}
\newcommand{\RR}{\mathrm{R}}
\newcommand{\U}{\mathrm{U}}
\newcommand{\SU}{\mathrm{SU}}
\newcommand{\Br}{\mathop{}\!\mathrm{Br}}
\newcommand{\andeq}{\quad \mathrm{and} \quad}

\newdimen\tdim  
\tdim=\unitlength  
\def\bar{\overline}

\begin{document}

{\title{ Scalar Democracy}

\author{Christopher T. Hill}\email{hill@fnal.gov}
\affiliation{Fermi National Accelerator Laboratory\\
P.O. Box 500, Batavia, Illinois 60510, USA}
\author{Pedro A. N. Machado}\email{pmachado@fnal.gov}
\affiliation{Fermi National Accelerator Laboratory\\
P.O. Box 500, Batavia, Illinois 60510, USA}
\author{Anders E. Thomsen}\email{aethomsen@cp3.sdu.dk}
\affiliation{CP$^3$-Origins, University of Southern Denmark\\
Campusvej 55, DK-5230 Odense M, Denmark
}
\author{Jessica Turner}\email{jturner@fnal.gov}
\affiliation{Fermi National Accelerator Laboratory\\
P.O. Box 500, Batavia, Illinois 60510, USA}

\date{\today}

\begin{abstract}We conjecture that there exists a  scalar bound state for every pair of
fundamental fermions at a UV (``composite'') scale, $\Lambda\gg v_{\text{weak}}$.
This implies a large number of universally coupled, sub-critical
Higgs doublets. All but the Standard Model   Higgs are ``dormant,'' with large positive
squared masses and each receives a  small vacuum expectation values  via mixing with the Standard Model Higgs. 
Universal  couplings, modulo renormalization group running effects, flips the flavor problem 
into the
masses and mixings of the Higgs system. Doublets associated with
heavy fermion  masses, $b,c, \tau$ likely lie in the multi-TeV range, but may be observable at the current LHC,
or a high-luminosity and/or an energy-upgraded LHC.
In the lepton sector we are lead to a Higgs seesaw for neutrino masses, and corollary processes of 
observable  flavor violation. 
The observation of the first sequential doublet coupled to  $\bar{b}b$ with masses $\lesssim 3.5$ TeV
would lend credence to the hypothesis.
\end{abstract}

\preprint{\minibox[]{FERMILAB-PUB-19-068-T \\ CP3-Origins-2019-003 DNRF90}}
\maketitle

\section{Introduction}\label{sec:intro}

In the present paper we propose that every fermion pair binds to form a complex
scalar boson, due to a universal
attractive interaction at a very high scale, $\Lambda$.  Amongst many new states, including
lepto-quarks, colored isodoublets and singlets, etc., 
this hypothesis implies the existence of a large number of Higgs bosons.

We assume the lightest of the these is the Standard Model Higgs (SMH). 
The remaining doublets are sequentially heavier with positive
$M^2$'s, \ie,  ``dormant.''\footnote{We use the term ``dormant'' as distinct from ``inert'', which to us implies electroweak sterile scalars.} They will have universal
couplings to their constituent fermions at $\Lambda$, but renormalized couplings at the electroweak scale
that are $g\simeq 1$ for quarks, and $g_\ell \simeq 0.7$ for leptons.
Each Standard Model (SM) fermion acquires its mass through its coupling to the particular
Higgs doublet it comprises, which in turn mixes with the SMH to acquire a perturbative ``tadpole'' mass.  
In particular our present model
is ``subcritical:'' the negative $M^2$ of the SMH
arises from mixing with dormant Higgses.

The associated Higgs bosons thus become
lighter for the heavier fermions, and very heavy for the neutrinos. 
Though we have no theory of the masses and mixings of the large array
of composite scalars, we can make use of phenomenology. 
The model then becomes predictive, essentially because the Yukawa
couplings are determined. 
We call this system ``Scalar Democracy'' as it harkens back
to the ``Nuclear Democracy'' of the late 1960's. 

Sequential Higgs bosons have certainly been
considered previously, and we cannot review the vast literature.
Nonetheless, few theorists venture beyond a few Higgs bosons.
Many Higgs bosons arise
in the context of an extended gravity, such as a 
scheme described to us by  Bjorken~\cite{Bjorken}, which first inspired  the idea of Scalar Democracy. A model that presages some of our present discussion is the ``Private Higgs''~\cite{Porto:2007ed, BenTov:2012cx}.
A Coleman-Weinberg mechanism and Higgs portal interactions
lead one of us to a second Higgs doublet coupled to $\bar b b$ with $g=1$ at about $400$ GeV, (see \cite{Hill1}
and a long list of references therein). In this model,
the SMH is a $\bar{t}t$ composite state
\cite{Yama,BHL,HillSimmons}, but the predictions of minimal composite $\bar{t}t$ are
significantly modified by mixing. 
We note that recent work attempting to construct an ``asymptotically safe''  UV completion of the SM  
arrives at a similar 
structure in the Higgs sector to ours \cite{Litim:2014uca,Bond:2017wut,Abel:2018fls},
but for different reasons. 
The phenomenology of many Higgs bosons is, to our knowledge, essentially unexplored.

In our particular scenario, we count 18 sequential colorless doublets in the quark sector, and 18 in the lepton sector.
In the quark sector this includes the
lightest doublet, $H_0$, associated with the top quark, $(\bar{t}_\LL, \, \bar{b}_\LL) \tilde{H}_0 t_\RR$ (where $ \tilde{H}_0 $ is the charge conjugated Higgs), which is identified with the SM
Higgs doublet. This establishes the universal quark sector Yukawa coupling to be $g=1$.
The quark masses and mixings are then determined by the spectrum of the 18 quark-sector Higgs doublets.  
We will see the Higgs doublet associated with the $b$-quark,  $g (\bar{t}_\LL, \, \bar{b}_\LL) H_b b_\RR $, is expected to
have a mass  $\lesssim 3.5$ TeV.  We also have  18 doublets in the charged lepton sector. With the exception of $H_\tau$ these tend to be much
heavier as lepton masses are small. This framework provides three alternatives for the neutrino mass generation.
Interestingly, neutrinos could be Dirac or Majorana via a type I or II seesaw mechanism.

Many of the scalars providing  quark and lepton masses are well beyond present collider reach.
The lighter ones, associated with the $b$ and $c$ quarks, as well as the $\tau$ lepton  may have thresholds in the sub-10-TeV range
and thus accessible to the LHC and future upgrades and higher energy machines. 
However, the heavier states may leave indirect imprints on
flavor-changing observables in the quark and lepton sectors.

Scalar Democracy for us is  new dynamics 
with subcritical compositeness due to a universal interaction
in the far UV.  
We thus blend a few key ideas from compositeness
and mainly emphasize  that the Yukawa
couplings are universal at $\Lambda$, subject only to renormalization
group evolution from $\Lambda$ to the weak scale.  
This, together with the input masses and mixings of
the Higgs bosons, flips the quark and lepton flavor problems away from the issue of 
understanding a fundamental Yukawa-coupling matrix. 
The  puzzle of fermion mass hierarchy becomes one of understanding and disentangling the multi-Higgs mass spectrum. 
This conversion of the flavor problem is an interesting exercise in its own right.

It should be emphasized that it is hard to understand the
small Yukawa couplings in the SM, such as $g_\mathrm{electron}\sim 10^{-6}$. This cannot be generated perturbatively from zero, owing to the chiral symmetry of the electron.
It is natural that $g_\mathrm{electron}\sim 1 $ in its coupling to a new Higgs $H_e$, but through mass mixing, the induced coupling to the SMH  becomes small. This is a key motivation for a scheme such as the one presented here.

As we have emphasized, the present scheme 
does not provide any explanation of
the multiple Higgs masses and mixings.  
We will not concern ourselves with the overall naturalness, and treat  symmetry-breaking effects as inherent in the Higgs mass terms.
This is analogous to ``soft-symmetry-breaking''  in chiral Lagrangians and resonance models of the 1960's.
We do impose fine-tuning constraints in the low-energy effective theory of  $H_b$,  where the mixing generates
the largest feedback on the SMH, $H_0$ (level repulsion),
and the negative $M^2$ of the SMH can arise from this effect.

It is our conclusion, at least at a first pass with various simplifying assumptions made along the way, that such a theory can exist.
New phenomena  may show up at a high luminosity and/or energy-doubled LHC, 
and certainly at a $ 100$ TeV collider. $H_b$ in the lightest mass limit $\sim 1$~TeV is already accessible
to the LHC, while it would be seen in the higher mass range $\sim 3 $ TeV in upgraded LHC runs.
In our opinion, a robust theoretical spectroscopy of multiple 
Higgs bosons offers a rationale for luminosity and energy upgrades of the LHC and future ultra energetic machines. 
If the upgraded LHC were to fail to discover a pair of isodoublet Higgs bosons with universal coupling, 
such as the lightest $H_b$ or $H_\tau$ states in our model, then this scheme would be disfavored.

Analyzing, at least schematically, the phenomenological consequences and constraints of the Scalar Democracy hypothesis is our main goal. 
We begin in  \secref{sec:theory_motivation} with a theoretical ``motivation'' for this perspective. 
Then in \secref{sec:EFT}, we  summarize the dynamics of our model at low energies 
and arrive at a fairly simple effective Lagrangian describing the couplings to quarks and leptons of the multi-Higgs spectrum.
This is followed by a more detailed discussion of phenomenology in the subsequent section, \secref{sec:pheno}. 
Here we discuss the main production channels and collider prospects for the dormant Higgses. 
We will also discuss the implications of the dormant Higgses on flavor physics and the resulting bounds 
on their masses. 

A reader interested in a summary of the observable features of the model, including neutrino masses, may skip directly to \secref{sec:pheno}.

\section{Theoretical motivation} \label{sec:theory_motivation}

\subsection{Are There Many Scalars in Nature?}\label{sec:manyscalars}
Gravity is a universal attractive interaction.
All pairs of fundamental fermions must have attractive gravitational scattering amplitudes.
Near the Planck scale this may involve exotic, new strong dynamics, new condensates, instantons, \etc, 
and perhaps a new way to generate hierarchies. 

For example, enhanced gravitational interactions may trigger condensates, dynamically generating 
Majorana masses for the right-handed neutrinos \cite{Barenboim:2010db}. 
Similar effects may arise through gravitational instantons \cite{Dvali}.
Alternatively, intriguing extensions of gravity, such as brane-world models, extra dimensions~\cite{Randall}, 
or ``bigravity''~\cite{bigravity1,bigravity2}, have ingredients that may likewise produce universal attractive interactions at various scales. 
We can also generate a large subset of these scalars by postulating new, strong gauge dynamics, which is more 
concrete but will not be developed presently.

We  assume that a universal attractive interaction generates bound state scalar fields 
at a high energy scale $\Lambda$ (which may be of order $M_{\text{Planck}}$, but could be lower).
Our hypothesis is general and transcends a wide class of possible models. 
While we invoke a universal pairing force, such as gravity, this is nonetheless a schematic proposal.
However, when we tie this to the SMH it becomes
predictive.

We suppose the new scalar bosons are field-theoretic bound states of pairs of SM fermions. In analogy to the Nambu-Jona-Lasinio (NJL) model~\cite{Nambu}, the constituent fermions are free (unconfined) and any fermion pair couples with a common Yukawa coupling $g$ at scale $\Lambda$ to its bound-state scalar (the usual non-relativistic
intuition does not apply.). 
If the new interaction is medium-strong (subcritical), then bound state scalars will form with masses which will be below $\Lambda$ and positive. 
The formation of these bound states does not break any symmetries since a composite scalar inherits the quantum numbers of its constituents.  
Hence the scalars, in the absence of mixing effects, are presumably degenerate with  mass $M^2<\Lambda^2$ and cannot break flavor or gauge symmetries which would trigger proton decay, \etc

We do not have a theory of the scalar mass spectrum, in particular, and do not provide a  fine-tuning mechanism to generate a large hierarchy. 
Moreover, we require symmetry-breaking and mixing effects to
further empower the scenario. All of these effects will be considered to be ``soft-symmetry-breaking'' or ``relevant'' operators, and
simply inserted by hand. 

While the dynamics determining the spectrum is unknown, we know 
that it must respect the SM gauge symmetries. 
Rather than trying to concoct a theory of masses and mixing angles amongst a large number of 
Higgs doublets over a large range of scales, we will assume such  a theory exists and 
let phenomenology guide us.

We connect the hypothesis with the SM by postulating that the SMH is the lightest scalar doublet.
This would then  be the $ (\bar{t}_\LL,\, \bar{b}_\LL) t_\RR$ bound-state element of the  scalar complex~\cite{Yama,BHL}. 
The top-quark Yukawa coupling, $g$, is then the universal coupling for all {\em quark pairs} to their particular Higgs doublets.  
Leptons will likewise have a universal coupling to their Higgs bosons, $g_\ell$, but we expect that the renormalization group (RG) running will yield a ratio $g_{\ell}/g \sim 0.7$, due mainly to QCD (this is the analogue of the $\SU(5)$ relation for $m_b/m_\tau$ \cite{Georgi}).
In fact $g=1$ is close to the RG fixed point for the top quark~\cite{FP1,FP2}, which is not far from the prediction of compositeness~\cite{BHL}.
In fact, this scheme can bring the RG fixed point into concordance with $m_{t}=173$ GeV (see \equaref{eq:RGFP} and discussion below).

\subsection{Sketch of the Dynamics}\label{sec:dynamics}

The dynamics of the lowest-lying sequential Higgs bosons is rather simple.
Let us anticipate the dynamical implications in the case of the top and bottom quarks, 
and the three generations of neutrino masses.

The SMH, $ H_0 $, in isolation has the usual potential:
\beq\label{eq:potential0}
V_{\text{Higgs}}= -M_{0}^{2}H_0^{\dagger }H_0+\frac{\lambda}{2} (H_0^{\dagger }H_0)^2,
\eeq
where $-M_{0}^{2}$ is negative, and phenomenologically $
M_{0} \simeq 88$ GeV, $\lambda \simeq 1/4$ (the observed physical Higgs boson mass is $\sqrt{2} M_0 \simeq 125$ GeV).

New sequential Higgs doublets, $H_x$, are ``dormant,'' meaning they have the usual SMH 
electroweak quantum numbers, but owing to large, positive mass terms, $M_x^{2}H_x^{\dagger }H_x$, 
they do not directly undergo condensation. 
However, in order to generate the light quark and lepton masses, they must have small mixings to the SMH,
\bea\label{eq:potential}
 V&=& M_{H}^{2}H_0^{\dagger }H_0+\frac{\lambda}{2} (H_0^{\dagger }H_0)^2\nonumber \\
 & & +\sum_x \left( M_x^{2}H_x^{\dagger }H_x - \mu_x^{2}H_{0}^{\dagger }H_x \hc\right),
\eea
parametrized by $\mu_x$. 
The vacuum expectation values (VEV) of the dormant
Higgses are small, so the quartic terms should generally be negligible, and we ignore it.

The mass mixing causes each $H_x$ to acquire a small VEV (``tadpole'') of order 
\beq
\left\langle H_x \right\rangle = \binom{0}{\mu_x^2v/M_x^2},
\eeq
where $v\simeq 174$ GeV is the electroweak VEV.

Due to the mixing $H_{0}$ is then ``level-repelled'' down by an amount or order $\mu^4_x/M^2_x$: 
\beq
 -M_0^2= M_{H}^2-\sum_x \frac{ |\mu_x^{2}|^2}{M_x^{2}}.
\eeq
 We refer the interested reader to  Ref.~\cite{Dobrescu:2010mk} for a similar implementation of the mechanism in supersymmetric models.
Therefore, starting from a positive mass term, $M_{H}^{2}$, in \equaref{eq:potential}, this cumulative effect can explain why the SMH boson has the tachyonic, or negative, $-M_{0}^2 $, in \equaref{eq:potential0}.

Hence, in this  scenario there is only a single condensate, associated with a conventional ``Mexican hat potential,"  \ie, the SMH. 
The rest of the sequential Higgs bosons remain approximately pure doublets acquiring small tadpoles via mixing, which
is generally suppressed by $1/M_x^2$.

Let us presently anticipate the main discussion and illustrate how this setup operates for the top-bottom sub-system.
There we have the Yukawa interactions:
\bea
\mathcal{L}_\mathrm{yuk} = -g (\bar{t}_\LL, \, \bar{b}_\LL) \tilde{H}_{0} t_{\RR} - g (\bar{t}_\LL, \, \bar{b}_\LL) H_{b} b_{\RR}.
\eea
The top mass determines the common Yukawa coupling to be $g= m_{t}/v \simeq 1$.

The $b$-quark then receives its mass from $H_{b}$.
By assuming mixing $M_b^{2} H_b^{\dagger} H_b - \mu_b^{2}H_{0}^{\dagger }H_b + \text{h.c.}$, we find  
the induced tadpole VEV  to be $\left\langle H_b \right\rangle =(0,\,  v_b)$ where $ v_b= v {\mu_b^{2}}/{M_b^{2}}$.
This implies that the $ b $-quark mass is
\beq
m_b = gv \frac{\mu_b^{2}}{M_b^{2}}=m_{t} \frac{\mu_b^{2}}{M_b^{2}},
\eeq
while the Higgs boson mass is\footnote{Here and throughout the paper we will use $ M_x $ 
to denote Higgs masses and $ m_x $ to denote fermion masses.}
\beq
-M_{0}^2 = M_H^2-\frac{\mu_b^{4}}{M_b^{2}}.
\eeq
In order to avoid fine-tuned cancellations between the two terms on the RHS of the above equation
we anticipate that 
\beq
\frac{\mu_b^2}{M_b} \lta 100 \; \makebox{GeV},
\eeq
providing us with the estimate for an upper ``naturalness''  bound on $M_b$:~\footnote{Note that,
if we assumed $M_H^2=0$, then we could generate all of the tachyonic Higgs mass from the
mixing with $H_b$ and we would {\em predict}: $M_b=(88\;\text{GeV})m_t/m_b = 3.36$ TeV.}
\beq
M_b \lta \frac{m_{t}}{m_b}\cdot 100 \;\text{GeV} \simeq  3.5\;\text{TeV}.
\eeq
This is not a firm upper bound, as it is based on a tuning argument. 
Another possible way of parametrizing the low scale fine-tuning would be to assume that all $\mu$ parameters are at the electroweak scale, for instance $\mu=100$~GeV. We will use these two tuning criteria later as benchmarks for comparison.
Furthermore,
we can certainly have a lighter $M_b$  with less fine-tuning.
We note that a mass $M_b \sim 380$ GeV was previously obtained in a scheme 
in which the $H_b$ with $g=1$ drives a Coleman-Weinberg potential for the SMH, and this has 
not to our knowledge been ruled out by the LHC~\cite{Hill1}.

We will find that this simple scheme is consistent with flavor-physics constraints.
Interestingly, flavor dynamics is not then a fundamental consequence of the Yukawa 
coupling matrices as in the SM, but rather 
the low-energy suppression of flavor-changing neutral currents (FCNCs) is a consequence 
of the heaviness of these states---we also find some natural limits in which the structure 
of the theory is simplified and in which the phenomenological constraints are easy to understand.

The lightest doublets beyond SMH are $H_b$ 
associated with the $ b $-quark, $H_\tau$ associated with $\tau $, and $H_c$ associated with charm (and possibly $H_{ct}$ and $H_{tc}$, depending on how the CKM matrix is generated, as we will see later).
Depending on mixing assumptions, $H_\tau$ could in principle be lighter than  $H_b $.
We thus encourage LHC experimentalists to consider searching for these objects,
and we discuss collider phenomenology in \secref{sec:colliderlimits} and \secref{sec:colliderdiscovery}.

At the other end of the spectrum, we have the physics of neutrinos and charged leptons. In the Scalar Democracy, there are three alternatives for the neutrino mass mechanism. For example,  where neutrinos are Dirac their masses are generated  like any other fermion in the present model.
The Yukawa couplings for 
the neutrinos are similar to those for quarks and leptons, and the appropriate neutrino mass terms are generated 
by mixing of the dormant and SM Higgses.  
We generate neutrino mixings and small 
neutrino masses by having ultra-large and positive $M^2$ for 
the neutrino-Higgs fields in the matrix, \eg, 
\beq
m_{\nu_\alpha} \sim g_\ell v\frac{\mu_\alpha^{2}}{M_{\nu_\alpha}^{2}}.
\eeq
For neutrino masses of order $10^{-2}$ eV we thus have $ M_{\nu_\alpha} \sim 10^{11}$ GeV, too heavy for any observable collider or flavor signature.
Notwithstanding, the Scalar Democracy scenario may also lead to Majorana neutrinos via type-I and type-II seesaw mechanisms. In the type-I seesaw case, a light sterile neutrino is predicted, with mass in the keV$-$GeV range. A general discussion of all these possibilities will be provided in \secref{sec:neutrino}.

\subsection{Scalar Democracy and Counting Scalars}\label{sec:scalardof}

If we assume one boundstate per fermion pair at some high
scale $\Lambda$, then we can count the number of composite scalars in the theory.

All of the SM matter fields can be represented by $48$ two-component left-handed spinors, $\psi _{A}^{i}$. 
This includes all the left-handed and anti-right-handed fermions.
We can collect these into a large global $\SU(48)\times \U(1) $ multiplet, corresponding to the global symmetry assuming that the new dynamics are blind to the SM gauge interactions.
We emphasize that this is a dynamical symmetry, and familiar GUT theories that contain only the SM fermions will be gauged subgroups of this $\SU(48)$.
Here the indices $\left( i,j\right) $ run over all the $48$ flavor, doublet, and color degrees of freedom of the SM fermions.

The most general scalar-field  bilinear interaction
we can construct of these fields is
\bea
\epsilon ^{AB}\psi _{A}^{i}\psi _{B}^{j}\Theta _{ij}\hc,
\eea
where $\Theta _{ij}$ transforms as the symmetric $ \mathbf{1176} $ representation of the $\SU(48) \times \U(1)$ 
(this is analogue to the sextet representation of $ \SU(3) $). 
The field $\Theta _{ij}$ contains many complex scalar fields with assorted quantum numbers, 
including baryon and lepton number, color, and weak charges.

To make contact with the SM fields, we consider the usual $24$ left-handed quarks and leptons, $\Psi _{\LL,i}$, and the $24$ right-handed counterparts, $\Psi_{\RR,\widehat{i}} $.
 The index $i$ now runs over the chiral $\SU(24)_{\LL}$ and $\widehat{i}$ over the chiral $\SU(24)_{\RR} $ subgroups of $ \SU(48) $. 
With this notation, we can construct three interactions with bilinear fermion fields,
\bea
\label{fields}
\Phi_{i\widehat{j}}\overline{\Psi }_{L}^{i}\Psi _{R}^{\widehat{j}}+\Omega
_{ij}\overline{\Psi }_{L}^{i}\Psi _{R}^{jC}+\widehat{\Omega }_{\widehat{ij}%
}\overline{\Psi }_{R}^{\widehat{i}}\Psi _{L}^{\widehat{j}C}+\text{h.c}.,
\eea
where $ \Phi_{i\widehat{j}} $ is the $ (\mathbf{24}_\LL,\, \mathbf{24}_\RR) $ complex scalar field with $24^{2}=576$ complex degrees of freedom. 
$\Omega $ and $ \widehat{\Omega } $ are the symmetric $\mathbf{300} $ representation of $\SU(24)_{\LL}$ and $\SU(24)_{\RR}$ respectively
\footnote{
Notation: If $\Psi _{\LL}^{j} =\frac{1-\gamma ^{5}}{2} \Psi ^{j}$ is a left-handed spinor transforming as a $ \mathbf{24} $ under $\SU(24)_{\LL}$, then $%
 (\Psi _{\LL}^{j})^{C} = i\gamma ^{2}\gamma ^{0} \frac{(1-\gamma
^{5})^{\ast }}{2} (\Psi ^{j})^{\ast } = \left( i\gamma ^{2}\gamma
^{0}\right) \frac{1-\gamma ^{5}}{2} \left( -i\gamma ^{2}\gamma ^{0}\right)
\Psi ^{jC}=  \frac{1+\gamma ^{5}}{2} \Psi ^{jC} = (\Psi ^{jC})_{\RR}\equiv
\Psi _{\RR}^{jC}$ (in the notation of Bjorken and Drell) transforms as a $%
\overline{\mathbf{24}} $ under $\SU(24)_{\LL}$
and has a $\frac{1+\gamma ^{5}}{2}$ projection. 
}. Thus we have $\Phi \left( 576\right) +\Omega \left( 300\right) +\widehat{\Omega} \left( 300\right) =\allowbreak \Theta \left( 1176\right)$, matching the degrees of freedom of $\Theta _{ij}$.
Here $\Omega _{ij}$ and $\widehat{\Omega }_{ij}$ are the analogues of Majorana masses and carry fermion number, while $\Phi $ contains fermion number neutral fields, such as Higgs fields, in addition to ($B-L$) leptoquark multiplets.

The $\Phi ,\Omega$ and $\widehat{\Omega }$ fields can be viewed as the ``composite fields'' arising from a NJL model effective description of the new forces.
Consider just the $\SU(24)_{\LL} \times \SU(24)_{\RR}\times \U(1) \times \U(1)_{A}$ invariant NJL model:
\bea
\label{eq:NJL}
- \frac{g^{2}}{M^{2}} (\overline{\Psi}_{\LL}^{i} \Psi _{\RR}^{j}) (\overline{\Psi}_{\RR,i} \Psi_{\LL,j}),
\eea
where the negative sign denotes an attractive interaction in the potential.
It should be noted that we can equally well write current-current (and tensor-tensor) interactions, mediated by 
heavy spin-1 bosons (or Pauli-Fierz spin-2 gravitons); these will generally contain scalar
channels and will Fierz rearrange to effectively reduce to \equaref{eq:NJL} with the attractive signs.
There also exist the possibility of the following NJL models:
\bea
-\frac{g^{2}}{M^{2}} (\overline{%
\Psi }_{\LL}^{i} \Psi_{\RR}^{Cj} ) (\overline{\Psi}_{\RR,i}^{C} \Psi_{\LL,j} ) \;\;
\makebox{or} \;\; \left( \RR \leftrightarrow \LL\right),
\eea
which lead to the composite bosons $\Omega $ and $\widehat{\Omega}$. 
Such universal master interactions may arise as sub-sectors of more
general gravitational scattering amplitudes with many other effects near the Planck scale, $M\sim M_{\text{Planck}}$, including gravitational instantons or, at lower energy scales, \eg, from a strong bigravity force.
\equaref{eq:NJL} by itself can therefore be a starting point
of a discussion of a dynamically generated extended Higgs boson spectrum.

The first step to solving an NJL theory would be to factorize the interaction of \equaref{eq:NJL} by introducing auxiliary scalar fields.
This leads to the equation we started with,  \equaref{fields}, where  $\Phi ,\Omega$ and $\widehat{\Omega} $ are auxiliary fields.
The universal interaction will bind fermion pairs into scalars that are bound states of ordinary quarks and leptons and will generate a plethora of Higgs doublets. 
These bound states will have a universal Yukawa coupling $g$ at the scale $M^{2}$.
Moreover, with $g$ taking on a near-but-subcritical value, these bound states will generally have large positive masses but can be tuned to be lighter than $M$.

Symmetry-breaking effects will be required to split the spectroscopy, including the SMH down to its observed negative mass term.
All other doublets remain heavy, but will mix. 
The problem of solving an NJL model in the large-$ N $ fermion loop approximation, or equivalently by the RG, is discussed in detail in
Refs.~\cite{BHL,HillSimmons}. 

If $ g $ is supercritical then some or all multiplets will acquire negative renormalized masses, $ M^{2}(\mu \rightarrow 0)<0 $ and the theory develops a vacuum instability.
For example, the  field  $\Phi ^{ij}$ with a supercritical coupling will generally condense into a diagonal VEV, $\left\langle \Phi
_{ij} \right\rangle = V \delta _{ij}$, and this would become a spontaneously broken $\Sigma $-model of $ \SU(24)_{\LL} \times \SU(24)_{\RR} \times \U(1) \times \U(1)_A $.
In this supercritical case, all the fermions would acquire large, diagonal constituent masses of order $gV$,
inconsistent with observation.

The structure we have just outlined, even if subcritical, will contain many composite Higgs doublets.
While we want to have a reasonably deep binding of these scalars, the system must be near-to-but-subcritical such that no large VEVs will form.  
We would then expect a positive, diagonal mass-squared matrix amongst the many composite scalar states and all fermions would be massless. However, the effect of mixing, \ie, off-diagonal mass terms, can arise from ``extended interactions'' in analogy to ``extended technicolor'' or latticized extra dimensions.

Exactly how the scalar mass spectrum is generated is beyond the  scope of our present discussion.
We will simply assume such a spectrum of masses and mixings between the bound 
state scalars  that allows for a light sector from the SMH  to multi-TeV scales exists and extends up to the highest scales.  
We  assume that  $\Omega _{ij}$, $\widehat{\Omega }_{ij}$ and the color-carrying weak doublets
have very large positive $M^{2}$ and therefore we will  ignore them. 

Let us examine the quantum numbers of the spectrum of states in the $\Phi ^{ij}$ system.
There are  $24\times 24=\allowbreak 576\ $ composite complex scalars, and these devolve into the  following states upon gauging the fermions:
\begin{itemize}
\item $ 9 \times (\mathbf{1}, \, \mathbf{2},\, \tfrac{1}{2}) \sim \bar{Q}_{\LL} U_{\RR} $; $ 3^2 \times 1\times 2 = 18$ complex degrees of freedom (DoFs),

\item $ 9 \times (\mathbf{1}, \, \mathbf{2},\, -\tfrac{1}{2}) \sim \bar{Q}_{\LL} D_{\RR} $; $ 3^2 \times 1\times 2 = 18$ complex DoFs,

\item $ 9 \times (\mathbf{1}, \, \mathbf{2},\, \tfrac{1}{2}) \sim \bar{L}_{\LL} N_{\RR} $ leptonic; $ 3^2 \times 1\times 2 = 18$ complex DoFs,

\item $ 9 \times (\mathbf{1}, \, \mathbf{2},\, -\tfrac{1}{2}) \sim \bar{L}_{\LL} E_{\RR} $ leptonic; $ 3^2 \times 1\times 2 = 18$ complex DoFs,

\item $ 9 \times (\mathbf{8},\, \mathbf{2},\, \pm \tfrac{1}{2}) \sim \bar{Q}_\LL \lambda^{a} U_\RR [D_\RR] $; $ 3^2 \times 8 \times 2 \times 2 = 288$ complex DoFs, 

\item $ 9 \times (\mathbf{3},\, \mathbf{2},\, \tfrac{1}{6} [-\tfrac{5}{6}] ) \sim \bar{L}_\LL  U_\RR [D_\RR] $; $ 3^2 \times 3 \times 2 \times 2 = 108$ complex DoFs,

\item $ 9 \times (\bar{\mathbf{3}},\, \mathbf{2},\, -\tfrac{1}{6} [-\tfrac{7}{6}] ) \sim \bar{Q}_\LL N_\RR [E_\RR] $; $ 3^2 \times 3 \times 2 \times 2 = 108$ complex DoFs,
\end{itemize}
where the brackets denote the SM quantum numbers. 
The first four entries in the above list are the $36$ Higgs doublets in the quark and lepton
sectors respectively.

If we consider the $18$ scalars in the quark sector, ignoring
their masses and EW charges, we will have a 
Yukawa interaction at the scale $\Lambda$ that is
$\SU(6)_\LL\times \SU(6)_\RR $ invariant  and of the form:
\bea
g\bar{\Psi}_\LL  \Sigma \Psi_\RR \hc,
\eea
where $\Psi=(u,d,c,s,t,b)$ and $\Sigma$ is a $6\times 6$ complex matrix composed of $18$ doublets.
The renormalization group equation for $g$ is then determined to be
\bea \label{eq:RGFP}
&&(16\pi^2)\frac{\mathrm{d} g}{\mathrm{d} \ln(\mu)} = g(9 g^2 - 8 g_3^2-\kappa)
\eea
at one-loop order, where $\kappa$ includes the smaller electroweak corrections
(which breaks the $\SU(6)_\LL \times \SU(6)_\RR $ invariance).
This describes the running of $g$ down to scales at which the various
Higgs doublets decouple.  Likewise, we have an equation for the
lepton sector, $g\rightarrow g_\ell$,  where the $-8g_3^2$ term is dropped.
Assuming  the Planck mass corresponds
to a Landau pole in $g$ and that
all Higgs bosons are active down to the electroweak scale, and then we derive $g\simeq 0.93$ and $g_\ell \simeq 0.71$
(this also leads to some ``fine-structure'' as the electroweak terms in $\kappa$
split degeneracy of between the Yukawa couplings for the up- and down-type quarks;
the full details of this are beyond the scope of this  paper).

This result implies a top quark mass of $\simeq 161$ GeV.  This is the prediction of 
the modified RG fixed point 
(equivalent to a focus point) 
of \cite{FP2}  including additional Higgs bosons, and
 represents a significant improvement over the original minimal top condensation
models \cite{BHL}. This prediction is robust with respect to the
precise values,  $\Lambda$ and $g(\Lambda)$.  
If we include the masses of the heavier Higgs bosons (as discussed below)
and decouple them at their thresholds, the prediction will increase and we expect
it to converge on the observed top quark mass. 

Grand unification is greatly complicated by such dynamics. 
We relegate such an investigation to future work.

\section{Low Energy Effective Theory}\label{sec:EFT}

The observed Higgs boson must reside amongst the color singlet $\bar{Q}_{\LL} q_{\RR}$ doublets. We proceed under the simplifying assumption that the SM Higgs doublet, $H_0$, can be identified with the doublet that couples to the fermionic combination of a top quark pair
\beq
(\bar{t}_\LL,\, \bar{b}_\LL) t_\RR \sim H_{0}.
\eeq
This is the unique logical choice, as it has the largest Yukawa coupling in the Standard Model and our theory dictates that all quarks will have this universal coupling. 
This therefore recovers in part the top-condensation models~\cite{Yama,BHL,HillSimmons}.

We further reduce the scope of the problem
by assuming all of the dormant Higgs doublets apart from the 36 color singlet $\bar{Q}_\LL q_\RR $ and $ \bar{L}_\LL \ell_\RR $ doublets  are arbitrarily heavy and therefore decoupled from the low-energy effective theory. Although the phenomenology of these other scalars could be very interesting, this framework does not provide any insight on their mass scale.  Besides, we will assume for concreteness that the neutrino mass mechanism is the same as for charged fermions, and comment on alternatives in \secref{sec:neutrino}.

At this point, it is convenient to resort to a common notation for the 
individual Higgs bosons. 
Below the scale $ \Lambda $, the effective theory we are considering 
is a modification of the SM where the Higgs sector has been replaced by
\begin{multline} \label{eq:new_Lagrangian}
\mathcal{L} \supset \abs{D_\mu H'}^2 - V(H') 
- g \, \bar{Q}_{\LL}^{\prime i} H_{ij}^{\prime d} D_{\RR}^{\prime j } - g \, \bar{Q}_{\LL}^{\prime i} \tilde{H}_{ij}^{\prime u} U_{\RR}^{\prime j} \\
- g_\ell \, \bar{L}_{\LL}^{\prime i} H_{ij}^{\prime e} E_{\RR}^{\prime j} - g_\ell \, \bar{L}_{\LL}^{\prime i} \tilde{H}_{ij}^{\prime \nu} N_{\RR }^{\prime j} \hc
\end{multline}
The generational indices of the fermions are labeled by indices $ i,j $ 
and the primes indicate that we are working in the gauge eigenbasis. 
The Higgs doublets are denoted by $ H^{\prime f}_{ij} $ in their flavor 
eigenbasis, where $ f=u,d,\nu,e $ represents the fermion type which acquire mass from the doublet and $ i,j $ the generations they couple to.
Each doublet has an upper charged component and a lower neutral component, 
$H_x=(h_x^+, h^0_x)$ and 
we employ the charge conjugation convention $ \tilde{H} = i\sigma_2 H^\ast $. 
All fermionic masses and mixings are due to the VEVs of the Higgs bosons. 
Universality implies that all quarks have a Yukawa coupling $g\simeq1$ and leptons  $g_\ell \simeq 0.7$.

\subsection{The Higgs Potential} \label{sec:potential}
Formally, we can define the 36 Higgs doublets as a ``vector,'' where we separate out the (mostly) SM-like Higgs:
$\left( H_{0}', H_a'\right)$.  
Thus, in addition to the SM-like Higgs, $H_{0}'$, we have 35 doublets
represented as, $ H'_a =\left( H_{1}',\ldots ,H_{35}' \right) $. Here, we define new notation  and apologize for that. Nevertheless, this index notation will only be applied in this subsection. In the remainder of the paper, we revert to the notation $H^f_{ij}$.

We may write out the multi-Higgs boson potential as 
	\begin{equation}
	\begin{split}
	V =\, & M_H^2 H_{0}^{\prime \dagger} H_{0}' + \frac{\lambda }{2}  \abs{H_0'}^4 + H^{\prime \dagger }_a M_{ab}^{2} H'_{b} \\
	& \quad - \left(H^{\prime \dagger}_{a} \mu^2_a H_0' \hc \right),
	\end{split}
	\end{equation}
where the positive quartic interactions of the dormant Higgses are taken to be negligible due to their large masses\footnote{A universal quartic coupling is another option for the quartic terms that confine Higgs mixing to the mass matrix.}. Thus the Higgs mixing is determined by the mass matrix. The mass terms that mix $H_a' $ with $H_0' $ can be viewed as vector, $\mu^{2}_a  = \left( \mu_{1}^{2},...,\mu_{35}^{2}\right) $, and $M^{2}_{ab}$ is a $35 \times 35$ hermitian mass matrix amongst the heavy scalars, which is taken to have positive eigenvalues.

It is important to note that there are {\em two distinct types of mixing} in this model sourced by $\mu^2_a $ and the off-diagonal part of $ M_{ab}^2 $ respectively. The former is responsible for mixing the SMH into all the Higgs flavor states, while the latter introduces mixing between the dormant Higgses. 
We will regard the off-diagonal elements, $\mu^2_{a}$, as perturbatively small when compared to $M_{ab}^2$.

The negligible quartic couplings leads to a mass degeneracy of the charged and neutral Higgs components of the dormant doublets.  Therefore, retaining the symmetry-breaking mass terms, the dormant Higgs fields will acquire tadpole VEVs via the $ \mu^2_a H^{\prime \dagger}_{a}  H_0' $ mixings, but their fluctuating fields will be degenerate doublets.

The Higgs mass terms are diagonalized at leading order (LO) in $ M_H^2, \mu_a^2/ M^2_{ab} $ by going to the basis defined by
	\begin{equation} \label{eq:Higgs_mass_eigenstates}
	\begin{split}
	H'_{a} &= H_a + M^{-2}_{ab} \mu_b^2 H_0 + \ldots \, , \\
	H_0' &= H_0 - \mu_a^{2\ast} M_{ab}^{-2} H_b + \ldots \, ,
	\end{split}
	\end{equation}
where $ H_0, H_a $ defines the physical Higgs doublets before EWSB. After rotating away the $\mu_a $ mixing, the equations of motion for the VEVs of the Higgs fields read
	\begin{equation}
	\begin{split}
	-M_0^2 \vev{H_0} + \lambda \abs{\vev{H_0}}^2 \vev{H_0} & =0, \\
	M^2_{ab} \vev{H_b} &=0,
	\end{split}
	\end{equation}
where 
	\begin{equation} \label{eq:higgs_mass_term}
	-M_0^2 = \left(M_H^2 - \mu^2_a M^{-2}_{ab} \mu^2_b \right) \simeq - \left(88\, \mathrm{GeV} \right)^2,
	\end{equation}
is the mass term for the SMH. The Higgs tadpoles are thus given by 
	\begin{equation}
	\vev{H_0} = \binom{0}{v} \andeq \vev{H_a} = \bm{0},
	\end{equation}
where $ v^2 = -M_0^2/\lambda \simeq (174\, \mathrm{GeV})^2 $ is the SM VEV. 

From our assumptions of the Higgs potential, we ignore further mixing between the massive Higgs bosons. The dormant Higgses constitute degenerate doublets and $ H_0 $ is directly identified with the SMH, its components being the physical Higgs and the 3 EW Goldstone Bosons. \equaref{eq:Higgs_mass_eigenstates} shows how $ H_0 $ is mixed into all the Higgs flavor states. Accordingly, 
the dormant Higgses acquire small VEVs which in turn produces the fermion masses in our model.

\subsection{Fermion masses and mixing}\label{sec:fermass}
The fermions acquire their masses from the Yukawa couplings as shown in \equaref{eq:new_Lagrangian}. In general, the mass matrices for the fermions are determined by the VEVs of the corresponding Higgs flavor eigenstates;
	\begin{equation} \label{eq:fermion_mass_matrices}
m_{ij}^{f} \equiv \left[\mathcal{L}_f^\dagger \, \mathrm{diag}(m_{1}^{f},\, m_{2}^{f},\, m_{3}^{f} ) \mathcal{R}_f \right]_{ij} = g_f \vev{h^{\prime f,0}_{ij}},
	\end{equation} 
where $g_f$ is the Higgs coupling to quarks ($g$) or leptons ($g_\ell$),  $ \mathcal{L}_f, \mathcal{R}_f $ are unitary matrices which rotate left- and right-handed fermions respectively from the gauge to the mass basis. We will proceed to determine the physical couplings between the fermions and the new Higgses assuming the Higgs potential outlined in the previous section and furthermore take the limit where the mass matrix, $M_{ab}^{2}$, of the dormant Higgses is diagonal thereby eliminating dormant-Higgs mixing.  

First, we focus on the quarks whose mass eigenstates are given by 
	\begin{equation}
	\begin{split}
			U_{\LL} = \begin{pmatrix} u_\LL \\ c_\LL \\ t_\LL \end{pmatrix}  = \mathcal{L}_u Q^{\prime u}_\LL, \quad U_{\RR} = \begin{pmatrix} u_\RR \\ c_\RR \\ t_\RR \end{pmatrix}  = \mathcal{R}_u U'_\RR,\\
	D_{\LL} = \begin{pmatrix} d_\LL \\ s_\LL \\ b_\LL \end{pmatrix}  = \mathcal{L}_d Q^{\prime d}_\LL, \quad D_{\RR} = \begin{pmatrix} d_\RR \\ s_\RR \\ b_\RR \end{pmatrix}  = \mathcal{R}_d D'_\RR.
	\end{split} 
	\end{equation}
The couplings between the down-type quarks and the neutral components of the dormant Higgses, in the mass eigenbasis, are given by \footnote{For convenience we have identified \eg $ H_{23}^{d} = H_{sb} $. We apologize for the notational inefficiencies.}
	\begin{multline} \label{eq:eigenq}
		\mathcal{L} \supset - \bar{D}_\LL \left[ g\, \mathcal{L}_d \begin{pmatrix}
	h^0_{d} & h^0_{ds}& h^0_{db} \\
	h^0_{sd} & h^0_{s}& h^0_{sb} \\
	h^0_{bd} & h^0_{bs}& h^0_{b}
	\end{pmatrix} \mathcal{R}_d^{\dagger} \right. \\
	\quad  + \left. \begin{pmatrix}
	m_d & & \\ & m_s & \\ & & m_b
	\end{pmatrix} \left(1 + \dfrac{h}{v}\right) \right] D_\RR,
	\end{multline}
and likewise for  the up-type quarks 
	\begin{multline} \label{eq:up-quark_couplings}
	\mathcal{L} \supset - \bar{U}_\LL \left[ g\, \mathcal{L}_u \begin{pmatrix}
	h^0_{u} & h^0_{uc}& h^0_{ut} \\
	h^0_{cu} & h^0_{c}& h^0_{ct} \\
	h^0_{tu} & h^0_{tc}&  [h]
	\end{pmatrix} \mathcal{R}_u^{\dagger} \right. \\
	\quad  + \left. \begin{pmatrix}
	m_u & & \\ & m_c & \\ & & m_t
	\end{pmatrix} \left(1 + \dfrac{h}{v}\right) \right] U_\RR .
	\end{multline}	
	where $[h] = - \mu^{2\ast}_a M^{-2}_{ab} h_b^{0}$ arises from the feedback on the SMH. 
Similarly in the lepton sector we set the mass eigenstates to be  
	\begin{equation}
	\begin{split}
	N_{\LL} = \begin{pmatrix} \nu_{1\LL} \\ \nu_{2 \LL} \\ \nu_{3\LL} \end{pmatrix}  = \mathcal{L}_\nu L^{\prime \nu}_\LL, \quad N_{\RR} = \begin{pmatrix} \nu_{1\RR} \\ \nu_{2\RR} \\ \nu_{3\RR} \end{pmatrix}  = \mathcal{R}_\nu N'_\RR,\\
	E_{\LL} = \begin{pmatrix} e_\LL \\ \mu_\LL \\ \tau_\LL \end{pmatrix}  = \mathcal{L}_e L^{\prime e}_\LL, \quad E_{\RR} = \begin{pmatrix} e_\RR \\ \mu_\RR \\ \tau_\RR \end{pmatrix}  = \mathcal{R}_e E'_\RR,
	\end{split} 
	\end{equation}
which gives interactions with the neutral Higgs bosons of the following form:
	\begin{multline}
	\mathcal{L} \supset - \bar{E}_\LL \left[ g_\ell \, \mathcal{L}_e \begin{pmatrix}
	h^0_{e} & h^0_{e\mu}& h^0_{e\tau } \\
	h^0_{\mu e} & h^0_{\mu}& h^0_{\mu \tau} \\
	h^0_{\tau e} & h^0_{\tau \mu}& h^0_{\tau}
	\end{pmatrix} \mathcal{R}_e^{\dagger} \right. \\
	\quad + \left. \begin{pmatrix}
	m_e & & \\ & m_\mu & \\ & & m_\tau
	\end{pmatrix} \left(1 + \dfrac{h}{v}\right) \right] E_\RR
	\end{multline}
and 
	\begin{multline}\label{eq:eigenell}
	\mathcal{L} \supset - \bar{N}_\LL \left[ g_\ell \, \mathcal{L}_\nu \begin{pmatrix}
	h^0_{1} & h^0_{12}& h^0_{13 } \\
	h^0_{21} & h^0_{2}& h^0_{23} \\
	h^0_{31} & h^0_{32}& h^0_{3}
	\end{pmatrix} \mathcal{R}_\nu^{\dagger} \right. \\
	\quad + \left. \begin{pmatrix}
	m_1 & & \\ & m_2 & \\ & & m_3
	\end{pmatrix} \left(1 + \dfrac{h}{v}\right) \right] N_\RR.	
	\end{multline}
A similar construction for the coupling of the fermions to the charged Higgses follows straightforwardly. We observe that $ h $, the neutral component of $ H_0 $, is completely indistinguishable from the SMH.

From the previously outlined assumptions on the Higgs sector we are able 
to estimate the masses of the dormant Higgses. 
Using \equaref{eq:fermion_mass_matrices} and \equaref{eq:Higgs_mass_eigenstates}, along
 with diagonal dormant mass matrix, we find that the fermion mass matrix is related to the Higgs mass terms by
	\begin{equation}
	m^{f}_{ij} = \left(\dfrac{g_f }{g}\right) \dfrac{\mu_{ij}^{f\, 2} }{ M^{f\, 2}_{ij} }m_t.
	\end{equation} 
To avoid a large fine-tuning in the SMH mass term, we expect $ \mu^4 / M^2 \lesssim \left(100 \, \mathrm{GeV} \right)^2 $. We thus arrive at the estimate 
	\begin{equation} \label{eq:mass_estimate}
	M^{f}_{ij} \lesssim \left(\dfrac{g_f}{g}\right)\dfrac{ m_t}{m^{f}_{ij} }\cdot 100 \, \mathrm{GeV}.
	\end{equation}
We therefore expect the dormant Higgs masses to be inversely proportional to the corresponding entry in the fermion mass matrix. Once we have made an ansatz for the fermion rotation matrices, $ \mathcal{L}_f, \mathcal{R}_f $, this provides us with an estimate for the Higgs masses. The bound shown in \equaref{eq:mass_estimate} should merely be viewed as a guideline, as it is based on a fine-tuning argument.  

Before we proceed to review two important limiting cases in the next section, we first define our notation: the Cabibbo-Kobayashi-Maskawa (CKM) matrix is given by $V_{\text{CKM}} = \mathcal{L}_{u} \mathcal{L}^\dagger_{d}$, and likewise the Pontecorvo–Maki–Nakagawa–Sakata (PMNS) matrix is given by $V_{\text{PMNS}} = \mathcal{L}_{\nu} \mathcal{L}^\dagger_{e} $.
In the limit of negligible dormant Higgs mixing, the required input to determine the Yukawa couplings is the set of mass rotations, $\mathcal{L}_f$ and $\mathcal{R}_f$. These rotation matrices are implicitly generated by the unknown $\mu_a^2$ which gives rise to the VEV structure we have encoded into the above expressions. 
Examining \equasref{eq:eigenq}{eq:eigenell}, we find  that the individual  matrices $\mathcal{L}_f, \mathcal{R}_{f}$ are unobservable in the Standard Model. However,  their combinations form the 
mixing matrices,  $ V_\mathrm{CKM} $ and $ V_{\mathrm{PMNS}}$, which enter  in the charged currents of the quark and lepton sectors respectively. 
Given sufficient dormant Higgs data, the structure of $\mathcal{L}$ and $\mathcal{R}$ could be determined, but at present  cannot be determined directly. We must therefore make an ansatz for their form.

\subsection{$ \mathcal{R}_f = \mathds{1}_3 $  and No-Mixing Limits} \label{sec:limit1}

We proceed by considering the various symmetries that arise in the special limit 
$\mathcal{R}_f\rightarrow \mathbb{1}_3$ and $m_q=0$. These are useful for relaxing 
rather stringent flavor mixing constraints.

Consider \equaref{eq:eigenq} with $\mathcal{R}_d=\mathbb{1}_3$ and the strange quark mass set to zero, $m_s=0$.  Furthermore, decompose the mass matrix of the dormant Higgs  fields, $H_a$, as a sum of a
diagonal $M_{\mathrm{diag}}^2$ and an off-diagonal $\delta M^{2}$ Hermitian
matrices:
\beq
M^{2}=M_{\mathrm{diag}}^2 +\delta M^{2}.
\eeq
In the case $\delta M^{2}=0$, we find there exists a discrete symmetry\footnote{This symmetry can only be approximately realized in a realistic part of the parameter space; it is broken by the quark mass terms and the Yukawa coupling between the top quark and the dormant Higgses.}, \eg, reflection of the right-handed strange quark, $s_R\rightarrow -s_R$
and the corresponding Higgs fields $(H_{bs},H_{s},H_{ds})\rightarrow -(H_{bs},H_{s},H_{ds})$.
This symmetry is a generalization of the Glashow-Weinberg symmetry \cite{WG}, and in our present case of many Higgs bosons this is restrictive.

First, we observe that in general $H_{s}$ will mediate an interaction of the form,
\beq
\frac{1}{M^2_{s}} (\bar{s}'_\LL s'_\RR) (\bar{s}'_\RR s'_\LL),
\eeq
in the gauge basis. 
In the case  $\mathcal{R}_d$ were not unity, this would contain mixed combinations such as
\begin{multline}
\frac{1}{M^2_{s}}|(\mathcal{L}_{d,22} \bar{s}_\LL+ \mathcal{L}_{d,12} \bar{d}_\LL  )(\mathcal{R}_{d,22}^\ast 
s_\RR+ \mathcal{R}_{d,12}^\ast {d}_\RR  ) |^2,
\\
\supset \frac{1}{M^2_{s}} \mathcal{L}_{d,22} \mathcal{L}_{d,12}\mathcal{R}_{d,22}^\ast\mathcal{R}_{d,12}^\ast (\bar{s}_\LL d_\RR) (\bar{s}_\RR d_\LL)+...
\end{multline}
in the flavor basis,
which results in  $\Delta S=2$ transitions. 
The $K_L$--$ K_S$ mass
splitting places severe limits on $M_{s}\gta 10^{3}$ TeV for left- and right-handed mixings of the order of the CKM matrix, while  
the mass estimate of  \equaref{eq:mass_estimate} suggests $ M_s \lesssim 100 $ TeV.
With the discrete symmetry, $\mathcal{R}_d = \mathbb{1}_3$, no such interactions are generated at tree-level, and the tension is substantially alleviated.
Moreover, the discrete symmetry forbids the similarly dangerous Higgs mixing term $\delta M_{ds,sd}^{2} H^\dagger_{ds} H_{sd}$, which mediates interactions such as $\frac{1}{M^2} (\bar{d}_\LL {s}_\RR) (\bar{d}_\RR s_\LL) $ directly.

\begin{table*}
\centering
\renewcommand{\arraystretch}{1.3}
\begin{tabular}{|p{3cm}p{4.1cm}p{3cm}l|} 
	\hline
	Higgs field & Fermion mass &  Case (1) [TeV]  &  Case (2) [TeV] \\
	\hline
	\hline
        $H^{\prime}_0 = v + \frac{h}{\sqrt{2}}$ & $m_t = gv=175$ GeV  & $m_H=0.125$  &  $m_H=0.125$ \\
	\hline
	$H^{\prime}_{b}=v\frac{\mu^2}{M_b^2} + H_{b}$ & $ m_b= gv\frac{\mu^2}{M_b^2}=4.5$ GeV   & $M_b=3.9$  &  $M_b=0.620  $     \\
	\hline
	$H^{\prime}_{\tau}=v\frac{\mu^2}{M_\tau^2} + H_{\tau}$ &  $ m_\tau= g_\ell v\frac{\mu^2}{M_\tau^2}=1.8$
	GeV &  $M_\tau= 6.8 $   &  $M_\tau= 0.825 $  \\
	\hline
	$H^{\prime}_{c}=v\frac{\mu^2}{M_b^2} + H_{c}$ &  $ m_c= gv\frac{\mu^2}{M_c^2}=1.3$ GeV &  $M_c=13.5$  &  $M_c=1.2$  \\
	\hline
	$H^{\prime}_{\mu}= v\frac{\mu^2}{M_\mu^2} + H_{\mu}$ &  $ m_\mu= g_\ell v\frac{\mu^2}{M_\mu^2}
	=106$ MeV &  $M_\mu= 1.2\times 10^2$  & $M_\mu= 3.4$   \\
	\hline
	$H^{\prime}_{s}=v\frac{\mu^2}{M_s^2} + H_{s} $ & $m_s= gv\frac{\mu^2}{M_s^2}=95$ MeV & $M_s= 1.8\times 10^2$  & $M_s= 4.3 $  \\
	\hline
	$H^{\prime}_{d}=v\frac{\mu^2}{M_d^2} + H_{d} $ &  $m_d= gv\frac{\mu^2}{M_d^2}=4.8$ MeV 
	& $M_d=3.6\times 10^3  $ & $M_d=19$  \\
	\hline
	$H^{\prime}_{u}=v\frac{\mu^2}{M_u^2} + H_{u} $ &  $m_u= gv\frac{\mu^2}{M_u^2}=2.3$ MeV 
	& $M_u= 7.6\times 10^3 $  & $M_u= 27 $ \\
	\hline
	$H^{\prime}_{e}= v\frac{\mu^2}{M_e^2} + H_{e}$ &   $m_e= g_\ell v\frac{\mu^2}{M_e^2}=0.5$ MeV
	& $M_e=2.45\times 10^4$  &  $M_e=49$ \\
	\hline
\end{tabular}
\caption{The non-mixing estimates for heavy dormant Higgs bosons masses assuming 
(1) the level-repulsion feedback on the 
Higgs mass term is limited to $(100 \,\text{GeV})^2$  for each of the quarks and leptons, hence $M_q= (100$ GeV$)(m_t/m_q )$ and 
$M_\ell= (100$ GeV$)(g_{\ell}m_t/m_{\ell})$.
(2) $\mu=100$ GeV for all mixings, hence $M_q= \mu (m_{t}/ m_{q})^{1/2}$  and $M_\ell= \mu (m_{t}g_{\ell}/gm_{q})^{1/2}$.
Here  $g=1$, $g_\ell=0.7$ and $ v= 175 $ GeV.  The scalar spectrum relevant to the neutrino mass generation is discussed in more detail in Sec.~\ref{sec:neutrino}. 
\label{tab:Higgs_masses}
}
\label{tab:Higgs_mass_est}
\end{table*}

At the one-loop level the Higgs-Higgs box diagrams, with exchange of the full doublet $H_s$, produces the effective interaction
\beq
\frac{1}{32\pi^2} \frac{1}{M_s^2} \left( \bar{s}'_\LL \gamma_\mu s'_\LL \right) \left( \bar{s}'_\LL \gamma_\mu s'_\LL \right).
\eeq
This operator is not restricted by $\mathcal{R}_d = \mathds{1}_3 $,
but yields a $\Delta S=2$ operator for nontrivial left-handed rotations, $\mathcal{L}_q$,
\beq
\frac{1}{32\pi^2} \frac{1}{M_s^2} (\mathcal{L}_{d,22})^2 (\mathcal{L}_{d,12}^\ast)^2 \left( \bar{s}_\LL \gamma_\mu d_\LL \right) \left( \bar{s}_\LL \gamma_\mu d_\LL \right).
\eeq
Comparing this effective operator with kaon-mixing bounds on the left-handed current~\cite{Isidori:2013ez}, we arrive at the limit $M_s \gtrsim 60 $ TeV, which is compatible with the mass estimate.

Our model can thus generate
CKM mixing,  with multiple flavorful Higgs doublets
and yet present no large FCNC  (no tree-level $\mathcal{O}(1/M_x^2)$ operators) if we
make $ \mathcal{R} = \mathds{1}_3$.
Typical cases include small breaking effects, giving $\mathcal{R} \neq \mathds{1}_3 $,
and flavor physics remains an important probe in a system like this.

In getting a sense of the model,
it is useful to consider the limit in which  $\mathcal{L}_f, \mathcal{R}_{f}$
are set to unity with $\delta M_{ab}^2=0 $ which turns off the mixing among dormant Higgses at leading order. We still allow
non-zero $\mu^2_a$ terms that mix the dormant Higgses to the SMH.

Let us focus on the up-quark system c.f. \equaref{eq:up-quark_couplings}.
The top quark has  acquired its mass by
direct coupling to the SMH, which defines the universal quark Yukawa
coupling $g=1$.
 In this limit we can identify the fields that develop tadpole VEVs and give the light quark masses, \equaref{eq:Higgs_mass_eigenstates}:
\bea
H'_{c} = H_{c} + \frac{\mu^2_c}{M_c^2} H_0, \qquad m_c= \frac{\mu_c ^2}{M_c^2}v,
\nonumber \\
H'_{u} = H_{u} +\frac{\mu^2_u}{M_u^2} H_0,	\qquad m_u = \frac{\mu^2_u}{M_u^2}v,
\eea
and likewise for the $d$, $ e $ and $ \nu $ sectors. In this limit
we can compute the diagonal dormant Higgs masses.
All other fields $H_x$ have no VEVs in this limit.
The full set of dormant Higgs masses are tabulated in \tabref{tab:Higgs_masses} under two different case assumptions:
	\begin{enumerate}
		\item	The feedback on the Higgs mass is maximal,
		but limited to $(100\,\text{GeV})^2 $  for each
		fermion (thus choosing a larger value of $\mu$ for lighter quarks), 
		hence $M_q= (m_t/m_q ) \cdot 100$ GeV and $M_\ell= (g_{\ell}m_t/m_{\ell})\cdot 100$ GeV.	
		\item $\mu=100 $ GeV for each SMH mixing term, $M_q = (m_t/m_q )^{1/2}\cdot 100$ GeV and $M_\ell= (g_{\ell}m_t/m_{\ell})^{1/2}\cdot 100$~GeV, which makes only the lightest $H_b$ , $H_\tau$, $H_c$ have any significant feedback on the SMH mass.
	\end{enumerate}
Hence, in the zero mixing angle limit, it is easy to see how the flavor
problem is mapped into an inverted mass spectrum of Higgs doublets.

\section{Quark-sector Phenomenology}\label{sec:pheno}
In \secref{sec:colliderlimits}, we  begin by asking  if relevant current LHC searches can place limits on the, possibly
multi-TeV-scale, lightest dormant Higgses. We follow with a
discussion of  the prospects of discovery both at the LHC and at future colliders of higher luminosities and center of mass (COM) energies in \secref{sec:colliderdiscovery}. Throughout \secref{sec:colliderlimits} and \secref{sec:colliderdiscovery}, we calculate the leading-order (LO) production cross section of the NP (new physics) processes using Universal FeynRules
Output (UFO) format of \texttt{FeynRules}~\cite{Alloul:2013bka} implemented into  \texttt{MadGraph5\_aMC@NLO} \cite{Alwall:2014hca} and applying 
the  NNPDF2.3 PDF set \cite{Ball:2012cx}. We implement the five-flavor scheme in order to account 
for the intrinsic $b$-quark content of the proton. 
In addition, constraints on the dormant Higgs masses from meson mixing and dedicated searches for flavor change in the charged lepton sector are outlined in \secref{sec:mesonmixing} and \secref{sec:leptonmixing} respectively.

\subsection{Current Limits from LHC Searches}\label{sec:colliderlimits}
In this section, we investigate if current limits from the LHC can place meaningful
constraints on the masses of the Higgs spectrum of the Scalar Democracy. 
The dormant Higgs most accessible at the LHC, is the lightest 
(non-SM) doublet which couples to the $b$-quarks, namely $H_{b}=(h_b^+,h_b^0)$, where $h_b^0$ is a  complex field.  
Let us first focus on the neutral component, $h^0_{b}$.

For single $h^0_{b}$ production, the dominant contribution is
via $b$-quark fusion  as shown in  \figref{fig:GFHb}.
We remind the reader, that the coupling of the $ b $-quarks to $h^0_{b}$ is
$g =1$. This implies that $h_b^0$ will decay approximately to two $b$-quarks
with a branching ratio of one.  As a consequence, a simple comparison 
of  a search for new resonances decaying into jets containing $ b $-hadrons is particularly amenable
for us to investigate if the best suited current searches may be able to exclude the relevant (multi-TeV) mass regime of $h^0_{b}$.
The ATLAS analysis  we choose to recast corresponds to a  dataset collected  at $\sqrt{s}=13$
TeV and  searches for two $b$-tagged jets with an invariant mass ($m_{bb}$) in the  $0.57-5$ TeV range \cite{Aaboud:2018tqo}. In the ``low" ($0.6 \leq m_{bb} (\text{TeV}) < 1.25$) and ``high" ($1.25 \leq m_{bb} (\text{TeV})  \leq 5.0$)  invariant mass regimes, $24.3$ $\text{fb}^{-1}$ and $36.1$ $\text{fb}^{-1}$ of integrated luminosity were analyzed  respectively.
The primary aim of this analysis is to search for
$Z^\prime$ decays to $ b $-quarks. Although this search is designed to place limits on a vector boson, rather than a scalar,
the final states are the same in both cases so the difference in  acceptance between our model and that of \cite{Aaboud:2018tqo} should
be small.

\begin{figure}[t!]
	\centering
	\includegraphics[width=0.45\textwidth]{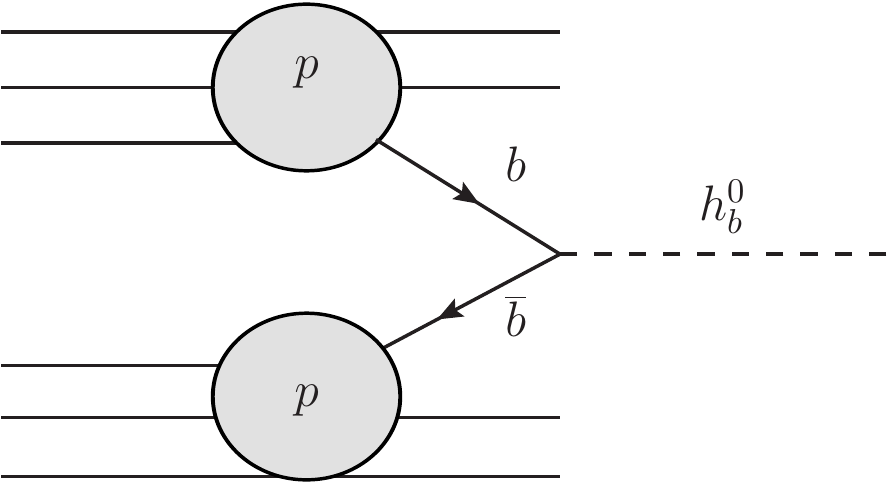}
	\caption{The dominant contribution to single $h^0_{b}$ production is $b$-quark fusion. $h^0_{b}$ decays with a branching ratio of approximately one to a two-$b$-quark final state.}
	\label{fig:GFHb}
\end{figure}

We note that 
single $h^0_{b}$ production is also possible via gluon fusion mediated by a bottom loop\footnote{The
top loop will also contribute to this process because of the mixing of the SMH with $h^0_{b}$. The modified coupling of the top-quarks to $h^0_{b}$  is approximately
$\simeq  -g m_{b}/m_{t}$, cf. \equaref{eq:up-quark_couplings}.}.  However, due to  the chiral suppression present in the
loop mediated processes, these contributions are subdominant to the tree-level production by more than two orders of magnitude  and therefore
will be ignored  presently.

The one- and two-sigma regions
for the expected upper limit on cross section of di-$b$-jet production is shown in \figref{fig:Hbbcollider}, as indicated by green and yellow respectively, while the observed upper limit, at $95\%\,\text{CL}$, is shown by the black circles.
There is no significant deviation between observed and expected limits. 
The details of the analysis used to calculate these curves can be found in the aforementioned reference.

 We have not performed a full analysis, including 
detector effects,  to calculate signal cross section and therefore the uncorrected
cross section will likely be a slight overestimate. 
To mitigate  this issue, the  LO  production cross section of $h^0_{b}$ was
 multiplied by a relatively stringent  efficiency factor, $\lvert \epsilon_{2b-\text{jet}}\rvert = 0.2$, to account for the two $b$-jet 
 reconstruction efficiency. 
    To justify this choice of efficiency, we refer to  \cite{Aaboud:2018tqo}, which details the $b$-tagging efficiencies for both 
``low"  and ``high"   mass regions. 
 They found  for the low-dijet-mass regions the efficiency for tagging two $b$-jets decreased from $0.5$ to $0.2$ as $m_{bb}$ was increased
 from $0.65$ to $1.25$ TeV. However, in the high-dijet-mass region, the  event  tagging efficiency for two $b$-jets  ranged from 
 $0.4-0.05$ for masses as $m_{bb}$ was increased
 from $1.25$ to $5.0$ TeV. 
\begin{figure}[t]
	\centering
	\includegraphics[width=0.45\textwidth]{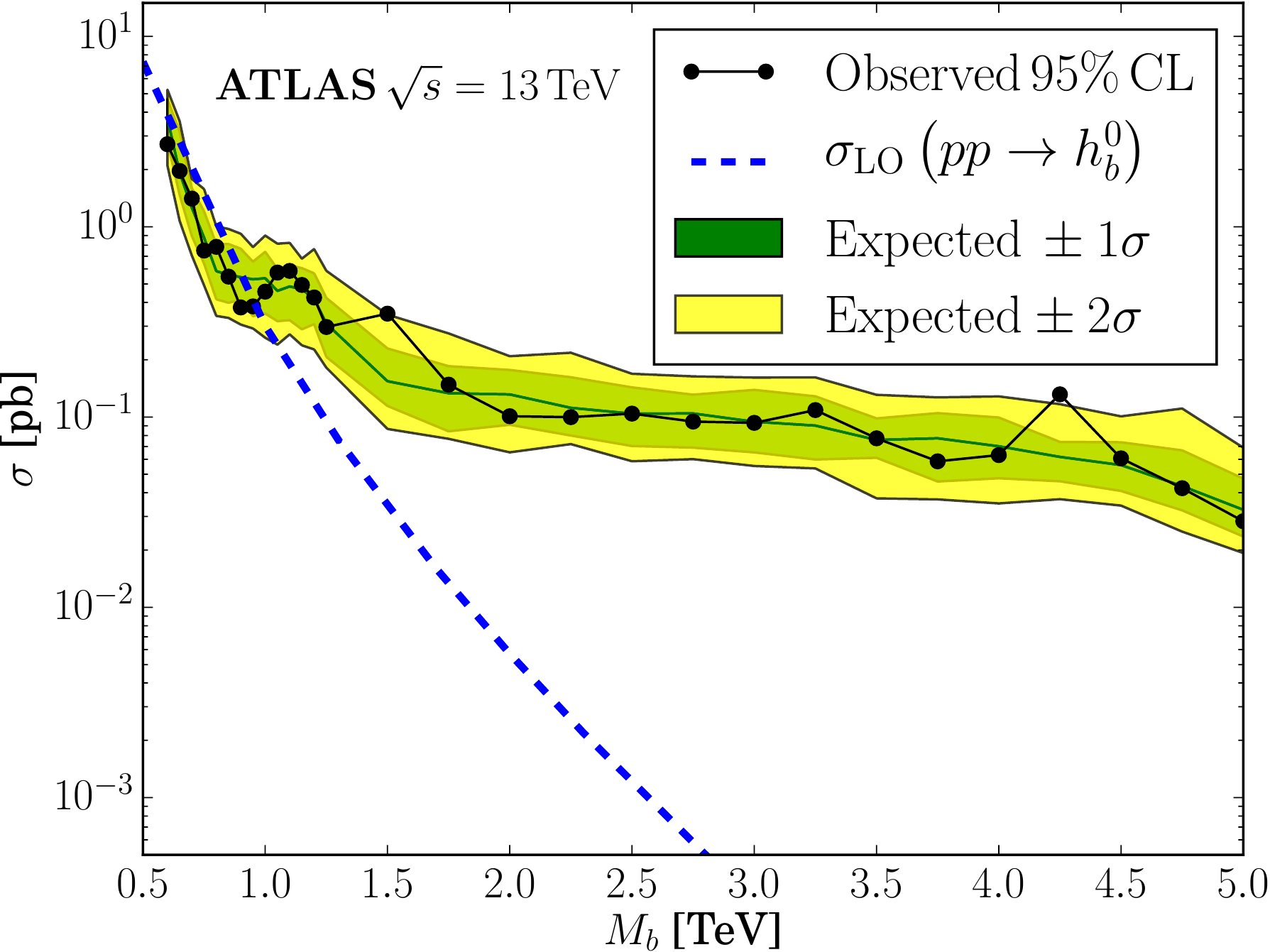}
	\caption{The green (yellow) region shows the $\pm 1\sigma$ ($\pm 2\sigma$) expected limit for  number of events from $Z^\prime$ decaying  to two $ b $-quarks. We note that the observed cross section (within 95$\%$ C.L) is 
within $2\sigma$ of the expected cross section. 
The blue lines shows the predicted LO inclusive cross section of $h^0_{b}$ production. The observed and expected data was analysed by an  ATLAS group \cite{Aaboud:2018tqo} and this data was made available 
via HEPData \cite{Maguire:2017ypu}.}
\label{fig:Hbbcollider}
\end{figure}
\begin{figure*}
	\centering
	\includegraphics[width=0.95\textwidth]{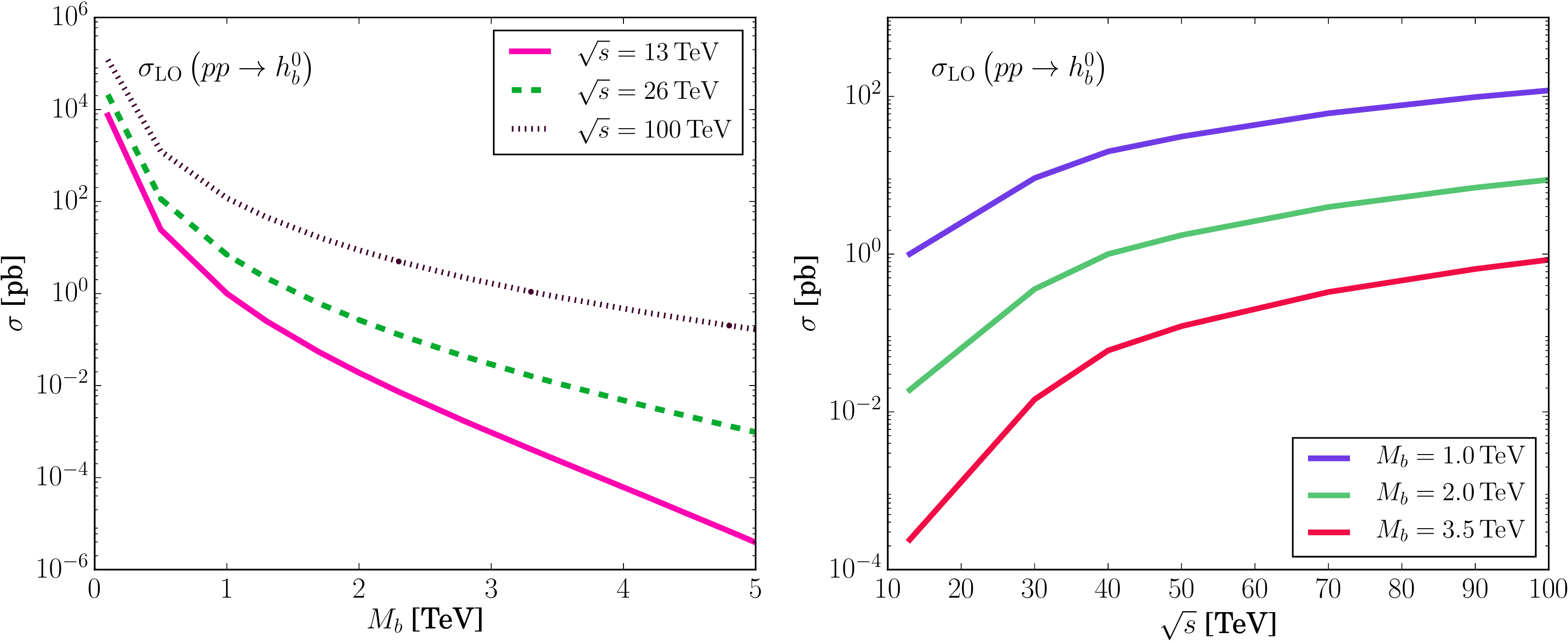}
	\caption{The left plot displays the total LO cross section  for single $h^0_{b}$ production  as a function of its mass, $M_b$, for three fixed COM energies $\sqrt{s} = 13, 26 $ and $100$ TeV as indicated by the solid fuchsia, dashed green and dotted purple lines.
	The right plot shows the LO cross section as a function of COM energy for three fixed mass $M_b$ = $1.0, 2.0, 3.5$ TeV as indicated by solid purple, green and red lines. }
\label{fig:Hbbcollider1}
\end{figure*}

The total inclusive cross section for charged and neutral Higgs production in heavy quark annihilation has been calculated at 
next-to-next-to-leading (NNLO) accuracy in  QCD \cite{Harlander:2015xur}.  For masses of new Higgses of $600$ GeV (which couple solely to $b$-quarks),
the  NNLO corrections are small, $\sigma_{\text{NNLO}}/\sigma_{\text{LO}} \sim 0.95$.  Moreover, it is likely this K-factor
will remain small for larger Higgs masses and therefore we did not  calculate higher order 
corrections to the cross section.
\begin{figure*}[t!]
	\centering
	\includegraphics[width=0.95\textwidth]{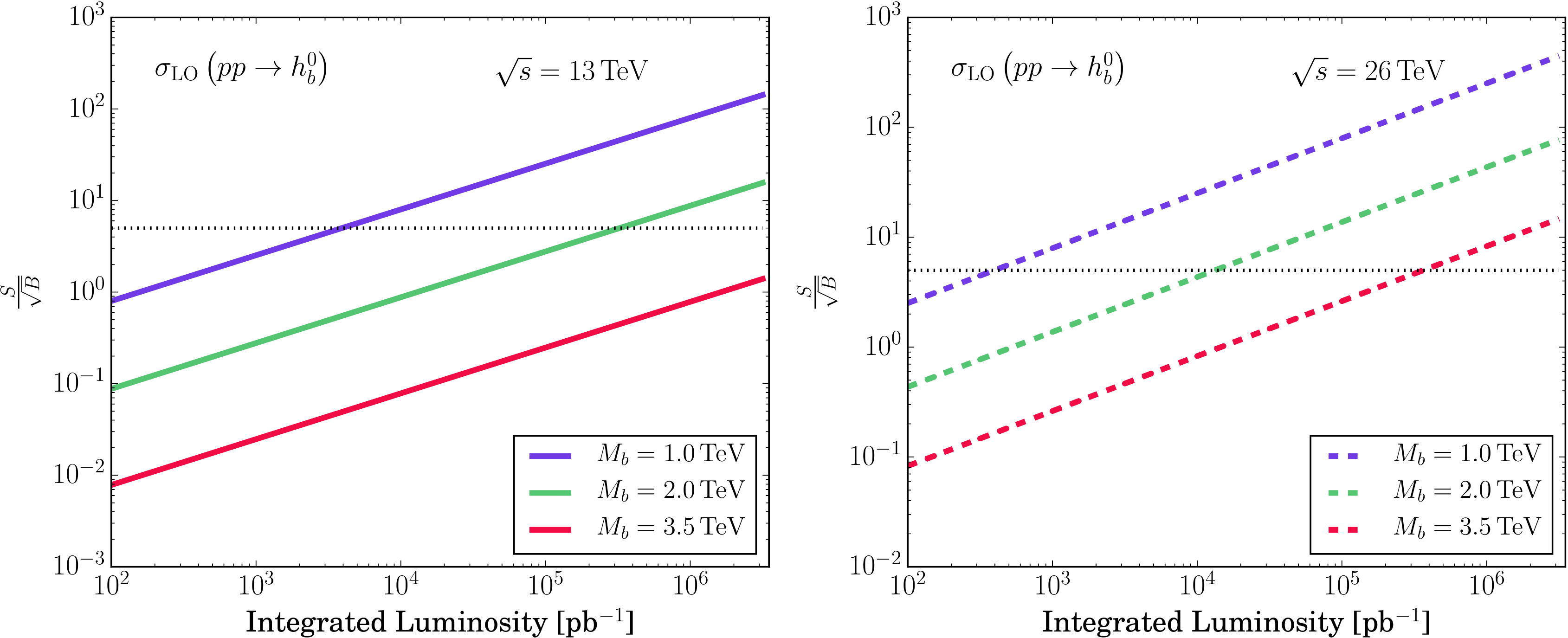}
	\caption{The left plot displays the significance, $S/\sqrt{B}$ as a function of integrated luminosity at $\sqrt{s}=13$ TeV for three fixed masses:
	$M_b = 1.0, 2.0, 3.5$ TeV as indicated by solid purple, green and red. The right plot shows the analogous information but with a COM energy, $\sqrt{s}=26$ TeV.}
\label{fig:Hbbcollider2}
\end{figure*}

Although, our reinterpretation lacks the sophistication of a full experimental analysis,
it allows for a rough comparison
of our signal cross section to a currently available dataset. From \figref{fig:Hbbcollider},
it seems likely that the LHC could exclude the
lower mass region, $M_b\lesssim 1.0$ TeV.

\subsection{Prospects for Future Discovery}\label{sec:colliderdiscovery}
The next most pertinent issue is the prospect for future discovery of
$h^0_{b}$ and the required   integrated luminosity.

In the left plot of  \figref{fig:Hbbcollider1}, we show the 
LO production cross section for single $h^0_b$ production  as a function of $h^0_b$ mass for three
COM energies: $13$, $26$ and $100$ TeV. As masses below $1$ TeV are disfavored
by data and the mass regime of $\sim3.5$ TeV is favored theoretically, we shall discuss the latter. 
For masses $M_b=3.5$ TeV, the corresponding cross section 
is $\sim2.4\times 10^{-4}$, $0.011$ and $2.0$ pb for $\sqrt{s} = 13, 26$ and $100$ TeV respectively. 
In the  right plot of \figref{fig:Hbbcollider1} we display the LO cross section as a function
of COM energy for three fixed masses: $M_b = 1.0, 2.0, 3.5$ TeV as shown in purple,
green and red respectively. 
Naturally, the cross section increases for higher COM energies. 

The most crucial
plots in addressing the question of observability are the two plots shown in \figref{fig:Hbbcollider2}.
The quantity of interest is the significance, defined as $S/\sqrt{B}$ where
$S$ is the expected number of signal counts and $B$ is the expected number of background (BG) counts. 
The  left and right plots display the significance, $S/\sqrt{B}$, as a function of integrated luminosity for
COM energies $\sqrt{s}=13, 26$ TeV respectively.
The significance is defined to be
\begin{equation}\label{eq:significance}
\frac{S}{\sqrt{B}} = 0.5(\mathcal{L}\,\lvert \epsilon_{2b-\text{jet}}\rvert)^{1/2} \frac{ \sigma_{\text{LO}}\left(p p \rightarrow h^0_{b}\right)}{\left[ \sigma_{\text{SM}}\left(p p \rightarrow b \overline{b} \right)\right]^{1/2}},
\end{equation}
where $\mathcal{L}$ is the integrated luminosity and $\lvert \epsilon_{2b-\text{jet}}\rvert=0.2$ is the applied reconstruction 
efficiency for two $b$-jets which we assume is the same for both signal and SM background.

In order to estimate the SM background, we applied a number of cuts on the
di-$b$-quark production:
\begin{itemize}
\item $p_{T}\left(b\right)\geq 100$ GeV
\item $M_b - \Gamma\!\left(h^0_b\right) \leq m_{bb}\leq M_b + \Gamma\!\left(h^0_b\right)$
\item $\lvert y \rvert >3.0$
\end{itemize}
where $p_{T}\left(b\right)$ is the transverse momentum
of each of the $b$-quarks,  $m_{bb}$ is the invariant mass of the $b$-quark system,  $\Gamma\left(h^0_b\right) = 3M_b/8\pi$ is the width of $h^0_b$ and $y$ is the rapidity.
We note that the cross sections have been calculated at the level of the hard matrix element.
The above cuts were applied to the invariant mass system of the two
$b$-jets  of the BG. We did not apply the cuts to the 
signal process but have chosen them in such a manner as to 
not significantly reduce the signal strength.
 In order to approximate the
effect of the cuts on the signal, we assumed a Breit-Wigner distribution for the differential
cross section as a function of the invariant mass
such that the chosen range captures approximately half of the signal events.
Subsequently, we multiplied the signal by a factor of 0.5, c.f \equaref{eq:significance}.

In both plots of  \figref{fig:Hbbcollider2}, the $S/\sqrt{B}$ value is shown 
for three fixed masses: $1.0$, $2.0$, $3.5$ TeV (as indicated by purple,
green and red colors respectively) and at two COM energies: $13$ and $26$ TeV (solid and dashed lines respectively).
The integrated luminosity takes the range between $\left(10^2-3.5\times10^{6}\right)\, \text{pb}^{-1}$.  
The value  $S/\sqrt{B} = 5$ is indicated by the dashed black lines and  provides an estimate of the required  integrated luminosity 
 for the discovery of $h^0_b$.

For current COM energies, the  masses, $ M_b = 1.0, 2.0$ TeV  would
be discoverable at integrated luminosities of $\sim2\times10^{3}$ and $2\times10^5$ $\text{pb}^{-1}$ respectively. 
However, for the heavier mass $ M_b = 3.5$ TeV, this would require greater than
$3.5\times10^{6}\,\text{pb}^{-1}$ of integrated luminosity.
With higher COM energies, such as $26$ TeV, the masses $M_b = 1.0, 2.0, 3.5$ TeV would all become
 accessible at $\sim 5\times10^2, 10^4,
2\times 10^{5}\, \text{pb}^{-1}$ respectively. 
Although we have not shown the projections of the significance for a COM collider with $\sqrt{s}=100$ TeV, we have found that $h^0_b$ of all three masses can be discovered with less than $3\,\text{ab}^{-1}$ of integrated luminosity.

In the event that $h^0_b$ were discovered,
a search and discovery of 
its charged counterpart, $h^\pm_{b}$,  would complete the doublet, $H_{b}$.
Although,  
$h^\pm_{b}$ is mass degenerate with $h^0_{b}$, its
production cross section is smaller by at least an order of magnitude,
 because
it  couples to $\overline{t}b$ or $\overline{b}t$ 
 and therefore cannot be produced from heavy quark fusion.
 The leading production channel of 
$h^{\pm}_b$ is in association with a bottom and a top quark, where the final state will be two $ b $-jets and two further jets.  Such a search would be more feasible than di-$h^{\pm}_{b}$ production,
which is mainly mediated via photons or off-shell Z-boson, as this process
 is kinematically suppressed
from the two Higgses in the final state.

One could also consider the production of
the lightest dormant doublet associated to the lepton sector, $H_{\tau}$. Due to its large coupling to $\tau$, $g_{\ell}=0.7$,  the neutral $h^0_{\tau}$ decays dominantly to $\tau\tau$. As $H_\tau$ does not couple strongly to quarks, it can only be produced via electroweak interactions, making it  challenging to probe at the LHC.
In spite of the fact the $\tau$ reconstruction efficiency is slightly higher than
that of $b$-quarks ($\lvert \epsilon_{\tau}\rvert \sim 0.8$ as measured at $\sqrt{s}=13$ TeV \cite{ATLAS:2017mpa}), as the 
 $h^0_{\tau}$ production is electroweak in nature, the cross section is significantly (several orders of magnitude) suppressed
compared with the analogous production of $h^0_{b}$.

Although this discussion has been largely schematic, and certainly
an in-depth analysis would be needed to search for the lightest dormant Higgs of the
Scalar Democracy, we find it an encouraging first step in demonstrating 
that the lightest dormant Higgs is within reach at current energies and most certainly future upgrades.

\subsection{Meson-antimeson mixing}\label{sec:mesonmixing}
Having discussed the main production channels for the dormant Higgses, we now turn to the implications of the Scalar Democracy for flavor phenomenology in the simple  limit discussed in \secref{sec:fermass}, where the dormant Higgses do not mix with each other. The NP parameters are reduced from several hundreds in the generic model down to a more manageable set consisting of the dormant Higgs masses and the fermion rotation matrices, $ \mathcal{L}_f, \mathcal{R}_f $, once we impose the requirement that SM fermion masses must be reproduced.  
 
Potentially large FCNCs may arise when the quark fields are rotated from their gauge to their mass eigenbasis and will place stringent constraints on the mass scale of these dormant Higgses. 
The constraints from flavor-violating processes in both the quark
and lepton sectors have been studied in depth the context of two Higgs doublet models, see e.g. \refref{Crivellin:2013wna}.

Integrating out the heavy Higgs mass eigenstates gives  NP contributions to the effective four-fermion operators at the low energy scale relevant for flavor physics. The neutral Higgses induce new four-fermion operators already at tree-level, contributing to the effective low-energy interactions among fermions with flavor indices ($q,p,r,s$), namely
	\begin{equation} \label{eq:eft_neutral}
	\mathcal{L}_\mathrm{eff} \supset - \hspace{-.5em} \sum_{f,ij \neq u,33} \! \dfrac{ g^2}{(M^{f}_{ij})^2 } \mathcal{L}_{f,qi} \mathcal{L}^{\ast}_{f,si} \mathcal{R}_{f,pj}^{\ast} \mathcal{R}_{f,rj} 
	(\bar{f}^{q}_\LL f_\RR^{p}) ( \bar{f}_\RR^{r} f_\LL^{s}).
	\end{equation}
The sum runs over all fermions except the $t$-quark denoted by ($u$,33), in order to avoid double counting, as this would be the SMH contribution. Besides, the SMH couplings are aligned with the fermion masses and thus do not lead to tree-level flavor changing transitions. The tree-level exchange is in most cases expected to be the leading NP contribution to the $ \Delta F=2 $ operators, as it is only suppressed by the mass of the dormant Higgses mediating the process. 

The right-handed rotation matrices, $ \mathcal{R}_f $, are not constrained a priori as they are unphysical in the SM. The limit $ \mathcal{R}_f = \mathds{1}_3 $ is thus viable and turn off the $ \Delta F=2 $ operators in \equaref{eq:eft_neutral}, as discussed in \secref{sec:limit1}. Near this limit the leading NP contributions will not come from tree-level contributions, but will rather be induced at the one-loop level from box diagrams with neutral or charged Higgses with or without SM charged currents $ W $ bosons. 
   
   \begin{figure*}
   	\centering
   	\includegraphics[width=.95\textwidth]{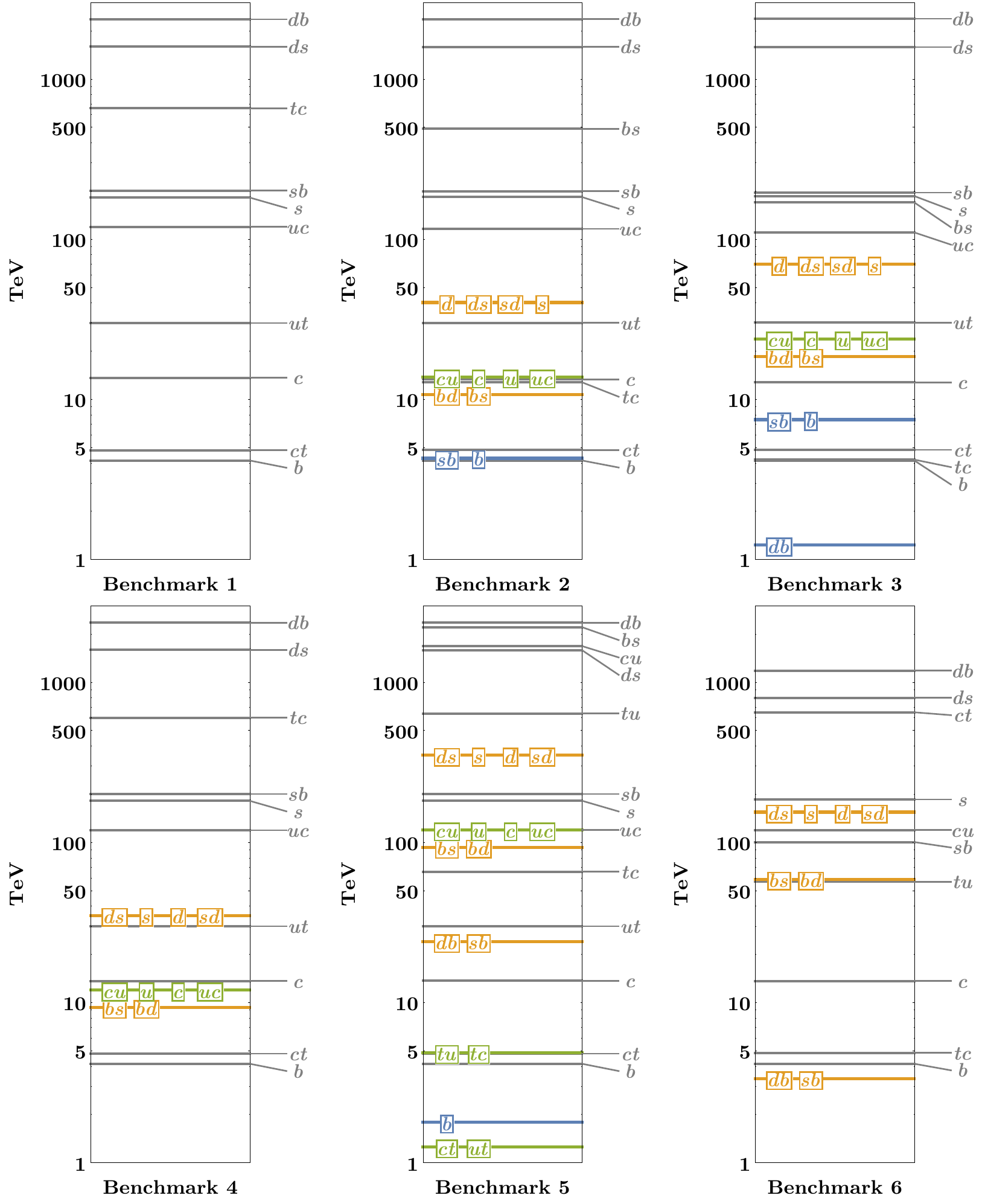}
   	\caption{Experimental constraints and mass estimates for the dormant Higgses in the quark sector in six different benchmarks (see text); the labels denote the indices of the corresponding Higgs. The gray lines are the mass estimates. The colored lines correspond to the most stringent experimental lower bound on each of the Higgs masses; orange if the constraint is from $ D_0 $ mixing, green from $ K_0 $ and blue from $ B_s $. If a mass-estimate entry is not shown, it is above the scale of the plot. Similarly if a mass bound is not shown it is below the scale of the plot. }
   	\label{fig:meson_constraints}
   \end{figure*}

We will focus on meson-antimeson mixing which are $ \Delta F = 2 $ observables that are loop and GIM suppressed in the SM. These observables are thus sensitive to NP degrees of freedom which propagate in the mixing amplitudes. As such, signals of NP may be constrained through comparison of observed meson mixing  and the SM predictions. 

Meson mixing have been observed in the $ D^0$--$\overline{D}^0 $, $ B_s $--$ \overline{B}_s $, $B^0 $--$ \overline{B}^0 $ and $ K^0 $--$ \overline{K}^0 $ systems. In each case, the observed mass splitting, $ \Delta M $, of the resulting meson mass eigenstates constrains the magnitude of the transition amplitudes. Although the experimental values are known to high precision \cite{Amhis:2014hma} the SM theory predictions are plagued by large hadronic uncertainties, and there is still room for NP ~\cite{Brod:2011ty,Lenz:2012mb}. The phases of the transition amplitudes give rise to CP-violation in the respective meson-antimeson systems as observed through the parameters $ \phi_D $, $ \epsilon_K $, $ S_{\psi K_s} $ and $ S_{\psi\phi} $. 

An exact analysis of the flavor constraints are somewhat beyond the scope of this paper. We will content ourselves with a rough estimate for the allowed mass range for each of the dormant Higgses. To this end, we apply the bounds of \citet{Isidori:2013ez} on the Wilson coefficients for the scalar exchange effective operators of Eq. \eqref{eq:eft_neutral}, cf. Table \ref{tab:meson_bounds}.  

To assess the parameter space, we will probe the NP constraints using a number of benchmark cases for the rotation matrices. In each case, flavor bounds will place a lower limit on the masses of the dormant Higgses, and we will compare these to the expected masses obtained from the fine-tuning arguments c.f. \equaref{eq:mass_estimate}. 

\begin{table}[t] \renewcommand{\arraystretch}{1.3}
\begin{tabular}{|c | c c|} 
\hline
& \multicolumn{2}{c|}{Bounds on $ C $ [$ \mathrm{TeV}^{-2} $]} \\ 
Operator & Re & Im
\\ \hline 
\hline 
$ C (\bar{s}_\RR d_\LL)(\bar{s}_\LL d_\RR) $ & $ 6.9\times 10^{-9} $ & $ 2.6\times 10^{-11} $\\
$ C (\bar{b}_\RR d_\LL)(\bar{b}_\LL d_\RR) $ & $ 3.9\times 10^{-7} $ & $ 1.7\times 10^{-7} $\\
$ C (\bar{b}_\RR s_\LL)(\bar{b}_\LL s_\RR) $ & $ 8.8\times 10^{-6} $ & $ 2.9\times 10^{-6} $\\
$ C (\bar{c}_\RR u_\LL)(\bar{c}_\LL u_\RR) $ & $ 5.7\times 10^{-8} $ & $ 1.1 \times 10^{-8} $ \\\hline
\end{tabular}
\caption{Bounds used to constrain the size of the NP contributions to FCNC operators relevant for meson-antimeson mixing~\cite{Isidori:2013ez}.}
\label{tab:meson_bounds}
\end{table}

The meson-antimeson mixing bounds on NP contribution to the $ \Delta F =2 $ operators, can be translated into bounds on the dormant Higgs masses once $ \mathcal{L}_{q}$ and  $\mathcal{R}_{q}$ are assumed. For each Higgs we determine the mass where it saturates each of the bounds in the absence of contribution from other dormant Higgses; the highest such mass is used as a lower bound for the corresponding Higgs mass. This is a somewhat simplified approach to the constraints, as it discounts possible interference between  contributions from multiple Higgses, not to mention the running of the effective operators below the EW scale. Since we are only exploring a small region of parameter space of the full model in any case, 
the bounds we obtain using these approximations remain a good indication for the scale where the Higgses become visible in flavor physics.

The benchmark points we consider are informed by the constraint that the left-handed quark mixings must reproduce the CKM matrix. Hence, it is unavoidable to have some left-handed mixing, albeit there is still freedom to put all of the left-handed mixing into either the up or the down sector. A good middle ground would seem to be taking a half and half approach. One will quickly find that with this assumption for the left-handed rotation, the meson-antimeson mixing constraints will favor small mixing of the right-handed quarks.
We consider the following benchmarks:\vspace{0.2cm}\\ 
{\bf Benchmarks}
    \begin{alignat}{2}  
	\text{(1)} \quad \mathcal{L}_{u} & = \mathcal{L}_{d}^{\dagger} = V_\mathrm{CKM}^{1/2}, \quad & & \mathcal{R}_u^{\dagger} = \mathcal{R}_d =\mathds{1}_3 \no \\
	\text{(2)} \quad \mathcal{L}_{u} & = \mathcal{L}_{d}^{\dagger} = V_\mathrm{CKM}^{1/2}, \quad & & \mathcal{R}_u = \mathcal{R}_d = V(1,1) \no \\
	\text{(3)} \quad \mathcal{L}_{u} & = \mathcal{L}_{d}^{\dagger} = V_\mathrm{CKM}^{1/2}, \quad & & \mathcal{R}_u = \mathcal{R}_d = V(3,3) \no \\
	\text{(4)} \quad  \mathcal{L}_{u} & = \mathcal{L}_{d}^{\dagger} = V_\mathrm{CKM}^{1/2}, \quad  & &\mathcal{R}_u^{\dagger} = \mathcal{R}_d = V_\mathrm{CKM}^{1/3000}\no \\
	\text{(5)} \quad  \mathcal{L}_{u} & = \mathcal{L}_{d}^{\dagger} = V_\mathrm{CKM}^{1/2}, \quad  & &\mathcal{R}_u^{\dagger} = \mathcal{R}_d = V_\mathrm{CKM}^{1/30}\no \\
	\text{(6)} \quad \mathcal{L}_u &= \mathds{1}_3, && \mathcal{L}_d^{\dagger} = V_\mathrm{CKM}, \no\\
	\mathcal{R}_u^{\dagger} &= V_\mathrm{CKM}^{1/2}, && \mathcal{R}_d = V_\mathrm{CKM}^{1/300} \no
	\end{alignat}
where 
    \begin{equation}
    V(\eta_1, \eta_2) = \begin{pmatrix} 1 & \eta_1 \cdot 10^{-4} & 0 \\
    -\eta_1 \cdot 10^{-4} & 1 & 8 \eta_2 \cdot 10^{-3} \\ 
    0 & -8 \eta_2 \cdot 10^{-3} & 1  \end{pmatrix},
    \end{equation}
is an approximately unitary matrix. 
For each benchmark we have determined the lower bound individually for each dormant Higgs mass, for all the effective operators in \tabref{tab:meson_bounds} to satisfy their bounds. The resulting lower bounds are plotted in \figref{fig:meson_constraints}. As a comparison the gray lines in the figure show the expected masses of the Higgses assuming the absence of fine-tuning, cf. \equaref{eq:mass_estimate}. We proceed to investigate the benchmark points one at a time.

\emph{Benchmark 1} shows the mass spectrum of the theory in the absence of right-handed rotation. The Higgs mass estimates in \figref{fig:meson_constraints} exhibits the inverse hierarchy of the quark masses detailed in \tabref{tab:Higgs_masses} with the Higgses associated to the lighter quarks tending to be the heaviest. The non-trivial left-handed rotation will lower the mass estimates for the off-diagonal mass. In particular, $ H_{ct} $, must be almost as light as $H_b $ in order to generate a large off diagonal mass element in the fermion mass matrix. The corresponding plot in \figref{fig:meson_constraints} does not show any bounds on the Higgs masses, as we have not included the loop-induced effects here. 

\emph{Benchmarks 2 and 3} illustrate a marginal case, for how much right-handed rotation can be allowed without tension between mass estimates and flavor bounds. The matrix $V(\eta_1, \eta_2)$ used to parametrize the right-handed matrix is applied for both $\mathcal{R}_u$ and $\mathcal{R}_d $ for simplicity. The $(1,2)$ entry governs the mixing of the first two generations of quarks. As $\eta_1$ is increased as do the flavor bounds on $H_{u} $, $H_{uc} $, $H_{cu} $ and $H_{c} $, because they are governed by $ D^0$ mixing (between the $u$ and $c$ quarks). Similarly the bounds on the $H_{d} $, $H_{ds} $, $H_{sd} $ and $H_{s} $ from $ K^0$ are pushed up, though they are less problematic. The entry (1,3) of $V(\eta_1, \eta_2)$ governs the bound on $H_b$ from $B_s$ mixing in a similar manner. Going from the marginal case $\eta_1 = \eta_2 = 1$ in Benchmark 2, to the case $ \eta_1 = \eta_2 = 3 $ in Benchmark 3, we observe how all the corresponding flavor bounds are increased by a factor $\sqrt{3} $, giving a slight tension between mass estimates and flavor bounds. We would like to emphasize that this tension only exists in the complete absence of fine-tuning of the Higgs mass, $-M_0^2$. If one were to tolerate fine-tuning to the percent level, the Higgs mass estimates would increase by a factor 10.     

\emph{Benchmarks 4 and 5}  parametrize $\mathcal{R}_q$ in terms of powers of CKM matrix, to illustrate how large the right-handed mixings are allowed to be in terms of a more familiar matrix. Due to the relatively large mixing, $(V_\mathrm{CKM})_{12} \simeq (V_\mathrm{CKM})_{21} \simeq 0.23 $, one has to go to the case $ \mathcal{R}_u^{\dagger} = \mathcal{R}_d = V_\mathrm{CKM}^{1/3000} $ of Benchmark 4, before the flavor bounds decrease below the mass estimates of $H_c$. On the other hand, if one is willing to accept a percent-level fine tuning of the Higgs mass we can go to the case $ \mathcal{R}_u^{\dagger} = \mathcal{R}_d = V_\mathrm{CKM}^{1/30} $ of Benchmark 5. We stress that, the mass estimates shown in \figref{fig:meson_constraints} all are with the complete absence of fine-tuning. 

\emph{Benchmark 6} puts the entire  left-handed rotation into $\mathcal{L}_d $, and gives an alternative  example of evading the flavor bounds. In this case a large $\mathcal{R}_u $ can be tolerated without generating $D^0$ mixing at tree level. The main constraints in this case, is from $K^0$ mixing.   

Though our selection of a few benchmark points does not constitute an in-depth analysis of the flavor physics of the Scalar Democracy, it demonstrates that there is room for the framework to avoid tension with flavor physics. Yet, care has to be taken to ensure that one works in an allowed regions of parameter space.  

\section{Lepton-sector phenomenology}

\subsection{Neutrino sector}
\label{sec:neutrino}

In our framework, the mass spectrum of SM fermions is mapped onto a scalar mass spectrum where lighter scalars develop larger VEVs and thus leads to heavier fermion masses, see Eq.~(\ref{eq:mass_estimate}). 
As the masses of charged fermions have been precisely measured, a sharper prediction of the scalar spectrum can be made. 
We do not know the absolute value of neutrino masses, nor the mass generation mechanism, resulting in looser possibilities for the scalar bound states associated to neutrino masses. 
The goal of this section is to examine these possibilities.

Although the absolute values of neutrino masses are still to be measured, neutrino oscillations give us valuable information on their masses. From the KamLAND experiment and observations of the solar neutrino spectrum~\cite{Gando:2013nba, Abe:2016nxk}, the solar mass splitting~\footnote{The three neutrino mass eigenstates are labeled such that $\nu_1$ ($\nu_3$) has the largest (smallest) admixture of $\nu_e$.} has been measured to be $\Delta m^2_{21}\equiv m_2^2-m_1^2\simeq +7.4\times10^{-5}$~eV$^2$. The sign of $\Delta m^2_{21}$ is known due to matter potential effects on neutrino oscillations. Besides, atmospheric and accelerator neutrino oscillation experiments~\cite{Abe:2017aap, Adamson:2017qqn, Abe:2017uxa} are compatible with $|\Delta m^2_{31}|\equiv|m^2_3-m^2_1|\simeq 2.5\times 10^{-3}$~eV$^2$. Therefore, the following two neutrino mass orderings are still viable experimentally, $m_1<m_2<m_3$ or  $m_3<m_1<m_2$ (recent experiments   exhibit a slight preference towards the former scenario~\cite{Adamson:2017gxd, Abe:2017vif}). In addition, the sum of neutrino masses is bounded from above by $0.12$~eV from cosmological observations~\cite{Aghanim:2018eyx}. We can thus identify three extreme scenarios for the neutrino mass spectrum $(m_1,\,m_2,\,m_3)$: normal hierarchy $(0,\,0.008~{\rm eV},\,0.05~{\rm eV})$, inverted hierarchy $(0.05~{\rm eV},\,0.05~{\rm eV},\,0)$ and semi-degenerate neutrinos $\sim(0.03~{\rm eV},\,0.03~{\rm eV},\,0.06~{\rm eV})$. As the heaviest neutrino needs to have a mass above $0.05$~eV, that will typically translate into an upper bound for some of the scalars (see Table~\ref{tab:Higgs_masses_Neutrinos}).

\begin{table*}
\centering
\renewcommand{\arraystretch}{1.3}
\begin{tabular}{|c|p{3cm}p{3.6cm}p{4.1cm}l|} 
	\hline
	Scenario & Higgs field & Neutrino mass &  Case (1)  &  Case (2)  \\
	\hline
	\hline
	Dirac & $H_{\nu}=v\frac{\mu^2}{M_{\nu}^2} + H'_{\nu}$ & $m_{\nu} = g_\ell \frac{\mu^2}{M_{\nu}}<0.06$ eV &  $M_{\nu}>2\times10^{14}$ GeV 
	& $M_{\nu}>1.4\times 10^{8}$ GeV\\
	& & $m_{\nu}^{\rm heaviest} > 0.05$ eV &  $M_{\nu}^{\rm lightest}<2.4\times10^{14}$ GeV 
	& $M_{\nu}^{\rm lightest}<1.5\times 10^{8}$ GeV \\
	\hline
	Type-I & $S_{N}= \mu\frac{v^2}{M_{S_N}^2} + S'_{N}$ & $m_N=g_\ell \mu \frac{v^2}{M_{S_N}^2} $ &  
	No sharp prediction & see \figref{fig:neutrino} \\
	\hline 
	Type-II & $\Delta=\mu\frac{v^2}{M_{\Delta}^2} + \Delta'$ & $m_\nu=g_\ell \mu \frac{v^2}{M_\Delta^2}<0.06$ eV &  
	$M_{\Delta}> 2\times 10^{14}$ GeV & $M_{\Delta}> 1.8\times 10^8$ GeV \\
	& & $m_{\nu}^{\rm heaviest} > 0.05$ eV &  
	$M_\Delta^{\rm lightest}<2.4\times 10^{14}$ GeV  &  $M_{\Delta}^{\rm lightest}< 2\times 10^8$ GeV \\
	\hline 
\end{tabular}
\caption{
\label{tab:Higgs_masses_Neutrinos}
Dormant Higgs bosons masses, for different neutrino mass generation mechanisms, assuming (1) the level-repulsion feedback on the Higgs is limited to $(100~{\rm GeV})^2$ for each neutrino, and (2) $\mu=100$ GeV for all mixings (see text for details).
Here, $v=175$ GeV and $g_\ell=0.7$. The requirement of a minimum mass for the heaviest neutrino typically translates into a low bound on the mass of some scalar.}
\end{table*}

In the Scalar Democracy there are three viable alternatives for the generation of neutrino masses.
The neutrino mass generation can be similar to the charged fermion mass generation, as outlined in Eq.~(\ref{eq:eigenell}). In this case, neutrinos would be  Dirac fermions.
Nevertheless, due to the quantum numbers of the lepton doublet, $L_{\LL}$, and the right-handed neutrino, $N_{\RR}$, 
two other bound states can contribute to neutrino masses, namely, $S_N\sim \bar N_{\RR}N_{\LL}^C$, 
and $\Delta \sim \bar L_{\LL} L_{\RR}^C$. While $H_\nu$ may lead to Dirac neutrino masses, an accompanying non-zero 
expectation value for $S_N$ would constitute a Majorana mass term for right-handed neutrinos, realizing a type-I 
seesaw mechanism, whereas $\Delta$ may be identified as the scalar triplet in type-II seesaw models 
(see Refs.~\cite{Minkowski:1977sc, Mohapatra:1986bd, GellMann:1980vs, Mohapatra:1979ia, Mohapatra:1980yp, Yanagida:1979as, Schechter:1980gr, Lazarides:1980nt}).
Thus, the mechanism of neutrino masses depends on which dormant Higgses acquire a VEV. 

In the type-I seesaw realization, the relevant terms in the Lagrangian are (generation indices have been suppressed)
\begin{equation}
\mathcal{L}\supset -g_\ell \bar L_{\LL} \tilde H_\nu N_{\RR} -g_\ell S_N \bar N_{\RR}N_{\LL}^C,
\end{equation}
and hence neutrino masses are given by 
\begin{equation}
m_\nu^\text{type-I}  = g_\ell\frac{  \vev{H_\nu}^2}{\vev{S_N}},\qquad m_N   = g_\ell \vev{S_N}.
\end{equation}
Because $S_N$ is a SM singlet, it will not acquire its VEV from mixing with the SMH as in Eq.~(\ref{eq:potential}), 
but rather from the term $\mu_S H_0^\dagger H_0 S_N$, which gives, $\vev{S_N}=\mu_S v^2/M_S^2$.
Requiring electroweak values for the dimensionful parameter $\mu_S=100$~GeV, and imposing constraints on singlet-Higgs mixing and active-sterile neutrino mixing leads to the allowed region in the $(M_S,M_\nu)$ plane shown in Fig.~\ref{fig:neutrino}. The gray regions are excluded due to mixings while the blue region predicts too heavy neutrinos~\footnote{Note that the constraint on the active-sterile mixing strongly depends on the mass $m_N$ and the active flavor~\cite{deGouvea:2015euy}. The value chosen here is extremely conservative.}. The atmospheric mass splitting requires at least one neutrino to be heavier than $0.05$~eV. A  dashed black line corresponding to $m_\nu=0.01$~eV is drawn to guide the reader. Notice that a sort of seesaw mechanism between the sterile neutrino and the singlet scalar is in place: the heavier is $S$, the smaller  its VEV and thus the lighter is $N$ (see arrows in Fig.~\ref{fig:neutrino}). If instead we require the level-repulsion feedback to the SMH mass to be small, no sharp prediction can be made about $M_S$, as it can always be made arbitrarily large leading to pseudo-Dirac neutrinos.

\begin{figure}[t!]
	\centering
	\includegraphics[width=0.45\textwidth]{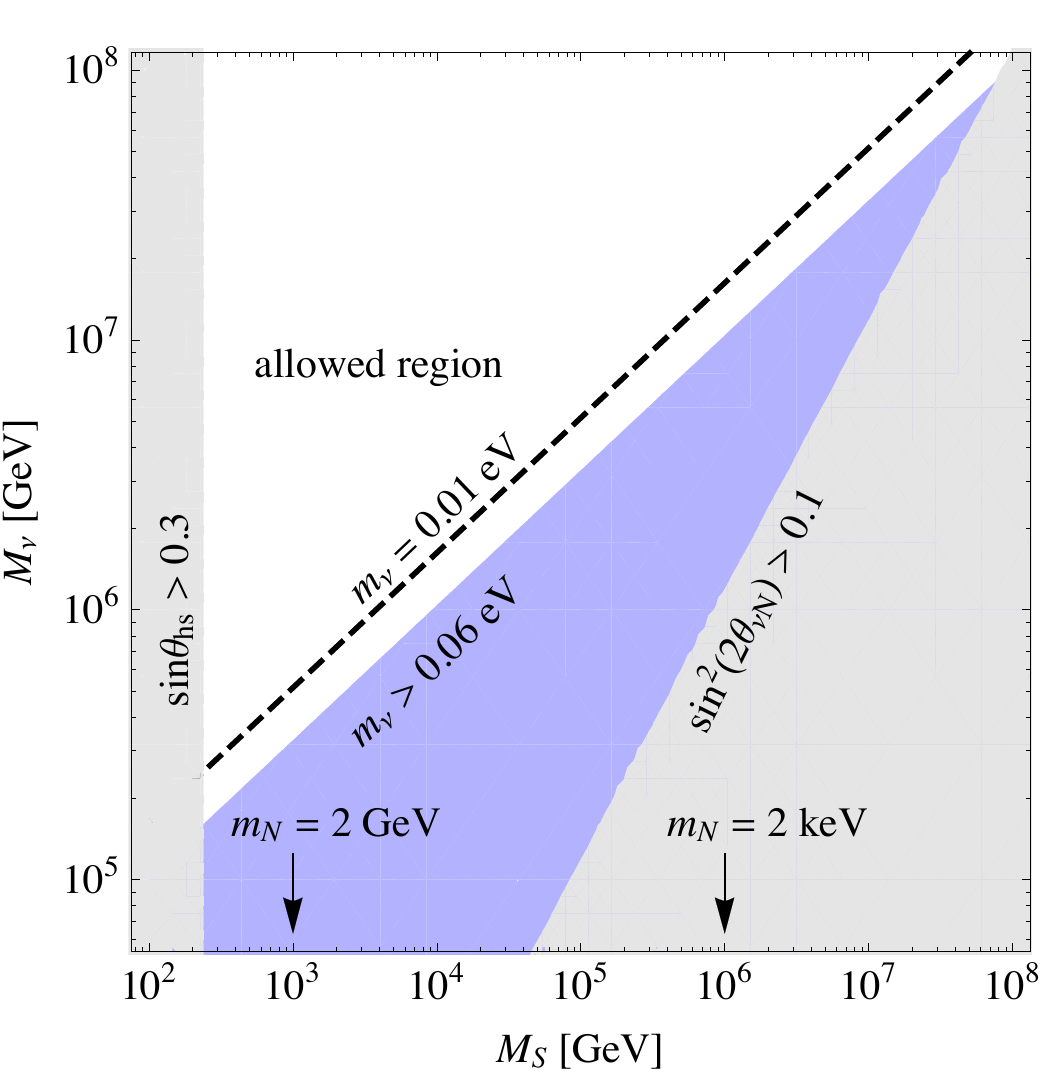}
	\caption{Mass of scalar singlet $S$ versus scalar doublet $H_\nu$ in the Scalar Democracy type-I seesaw scenario, under the assumption that $\mu_S=\mu_\nu=100$~GeV (see text for details). The gray regions are rule out by constraints on singlet-Higgs mixing (leftmost region) and active-sterile neutrino mixing (rightmost region). The blue region predicts too large neutrino masses and is excluded by cosmological observations. The heaviest neutrino is above 0.05~eV, and thus at least one pair of scalars should have masses between the dashed black line and the blue region. The mass of the sterile neutrino $m_N$ is also indicated for two values of $M_S$.
	\label{fig:neutrino}}
\end{figure}

In the case of type-II seesaw, in the absence of $N_{R}$, the $\Delta$ would get its VEV from the usual term $\mu_\Delta H_0^T i\sigma_2\Delta^\dagger H_0$. 
Neutrino masses would be given by 
\begin{equation}
m_\nu^\text{type-II}
=0.42~{\rm eV}\left(\frac{g_\ell}{0.7}\right)\left(\frac{\mu_\Delta}{100~{\rm GeV}}\right)\left(\frac{10^8~{\rm GeV}}{M_\Delta}\right)^2,
\end{equation} 
such that electroweak values of $\mu_\Delta$ would correspond to scalar triplets around the $10^8$~GeV scale, as can be seen in Table~\ref{tab:Higgs_masses_Neutrinos}, ``Case (2)'' column. Restricting the level repulsion feedback on the Higgs potential leads to exactly the same mass constraints as in the Dirac neutrino case. Obtaining scalar triplet masses accessible at the LHC, $M_\Delta\sim{\rm TeV}$ would 
require $\mu_\Delta\sim10~{\rm eV}$, as it happens in usual type-II seesaw scenarios.
For the three mechanisms discussed, a summary of the masses of the scalar bound states responsible for the neutrino spectrum can be found in Table~\ref{tab:Higgs_masses_Neutrinos} (labeled as ``Dirac'', ``Type-I'' and ``Type-II'').

\subsection{Lepton flavor violation}\label{sec:leptonmixing}
Any charged lepton flavor violating (CLFV) decay would immediately indicate the presence of new physics and as a consequence much effort has been put into detecting  such effects. The result of this work places severe constraints on the branching ratios for the different CLFV decays of the charged leptons. 

In the absence of mixing among dormant Higgses, the CLFV decays must be mediated by the  Higgses associated to the lepton mass generation.  As a result the CLFV transitions induced in our framework do not involve quarks. The relevant constraints are the radiative and the three-body  decays are shown in \tabref{tab:LFV3b}.

\begin{table}[h!] \renewcommand{\arraystretch}{1.3}
\centering
\begin{tabular}{| c | c  | }
\hline
  Process &   Experimental Upper Limit  \\
 \hline \hline
$\text{Br}\left(\mu^+\rightarrow e^+e^+e^- \right)$ 			& $1.0\times10^{-12}$ \cite{Bellgardt:1987du} \\
$\text{Br}\left(\tau^-\rightarrow \mu^-\mu^+\mu^- \right)$ 		& $2.1\times10^{-8}$ \cite{Hayasaka:2010np} \\
$\text{Br}\left(\tau^-\rightarrow e^-\mu^+\mu^- \right)$ 		& $2.7\times10^{-8}$ \cite{Hayasaka:2010np} \\
$\text{Br}\left(\tau^-\rightarrow \mu^- e^+ e^- \right)$ 			& $1.8\times10^{-8}$ \cite{Hayasaka:2010np} \\
$\text{Br}\left(\tau^-\rightarrow e^-e^-e^- \right)$ 			& $2.7\times10^{-8}$ \cite{Hayasaka:2010np} \\
$\text{Br}\left(\tau^-\rightarrow e^+\mu^-\mu^- \right)$		& $1.7\times10^{-8}$ \cite{Hayasaka:2010np} \\
$\text{Br}\left(\tau^-\rightarrow \mu^+e^-e^- \right)$ 			& $1.5\times10^{-8}$ \cite{Hayasaka:2010np} \\
$\text{Br}\left(\tau^\pm\rightarrow \mu^\pm \gamma \right)$        & $4.4\times 10^{-8}$ \cite{Hayasaka:2007vc} \\
$\text{Br}\left(\tau^\pm\rightarrow e^\pm \gamma \right)$		& $3.3\times 10^{-8}$ \cite{Hayasaka:2007vc} \\
$\text{Br}\left(\mu^+\rightarrow e^+ \gamma \right)$			 & $4.2\times 10^{-13}$ \cite{TheMEG:2016wtm} \\
\hline
\end{tabular}\caption{The  $90\%$ C.L. upper
limit on decay branching rates for LFV processes.}\label{tab:LFV3b}
\end{table}

The NP contribution to the decay of a charged lepton to another through the emission of a photon is governed, in the low-energy theory, by the dipole operator
	\begin{equation}
	\mathcal{L}_\mathrm{eff} \supset C_{ij} \overline{E}_{i\LL} \sigma^{\mu\nu} E_{j\RR} F^{\mu\nu} \hc,
	\end{equation}
where $i$ and $j$ are flavor indices and $C_{ij}$ has inverse mass dimension. The decay width is given by 
	\begin{equation}
	\Gamma(e_i \to e_j \gamma) = \dfrac{m_{e,i}^3}{4\pi} \left( \abs{C_{ij}}^2 + \abs{C_{ji}}^2\right).
	\end{equation}
In the Scalar Democracy, NP effects are induced by  diagrams where a dormant Higgs propagates in the loop.  The coefficients $ C_{ij} $ depend only on the rotation matrices of the external fermions and the Higgs mass.

For concreteness, we will assume that the Dirac neutrino mass mechanism is at work (see \tabref{tab:Higgs_masses_Neutrinos}). The NP contribution to the coefficients of the dipole operators, from both neutral and charged Higgses, is given by
	\begin{multline}
		C_{ij} = \dfrac{-e g_{\ell}^2}{384 \pi^2} \sum_{kl} \Bigg( \mathcal{L}_{e,ik} \mathcal{L}_{e,jk}^{\ast} \left[\dfrac{2 m_{e, j} }{(M^{e}_{kl})^2} - \dfrac{ m_{e, j} }{ (M^{\nu}_{kl})^2} \right]   \\
	\qquad \qquad + \mathcal{R}_{e,ik} \mathcal{R}^\ast_{e,jk} \dfrac{ m_{e,i}}{(M^{e}_{lk})^2} \Bigg),
	\end{multline}
to leading order in $ m_e^2/(M^e)^2 $, where $M^{e}_{kl}$  denotes the mass of the scalar that generates the $(kl)$ entry of the corresponding leptonic mass matrix. Due to the smallness of neutrino masses, the neutrino Higgs contribution is completely irrelevant to the radiative decay. As an aside, we note that a quick estimate of the NP contribution to $\mu\to\mu\gamma$ will show that even in the most optimistic scenario $C_{22}$ is several orders of magnitude too small to significantly influence the SM prediction for the muon $ g-2 $.

The charged lepton decay to three lighter charged leptons is mediated at tree-level by neutral Higgses giving the effective operators of \equaref{eq:eft_neutral}. The resulting decay width is
\begin{widetext}
	\begin{multline}
	\Gamma(e_p^{-} \to e_r^- e_s^+ e_t^-) \simeq \dfrac{ 7 g_{\ell}^4 m_{e,p}^5 }{6144 \pi^3}  
	\left( \abs{\sum_{ij} \dfrac{ \mathcal{L}_{e, r i} \mathcal{L}_{e, s i}^{\ast} \mathcal{R}_{e,pj}^{\ast} \mathcal{R}_{e,tj} }{(M^{e}_{ij})^2} }^2 + \abs{\sum_{ij} \dfrac{\mathcal{L}_{e, r i} \mathcal{L}_{e, p i}^{\ast} \mathcal{R}_{e,sj}^{\ast} \mathcal{R}_{e,tj} }{(M^{e}_{ij})^2} }^2 
	+ (r \longleftrightarrow t )\right).
	\end{multline}
	\end{widetext}
Many of these decays only violate flavor by one unit, and so, in contrast to the NP contribution to meson-antimeson mixing, the operator does not vanish for all dormant Higgses in the limit where $ \mathcal{R}_e = \mathds{1}_3 $.

\begin{figure*}
	\centering
	\includegraphics[width=.85\textwidth]{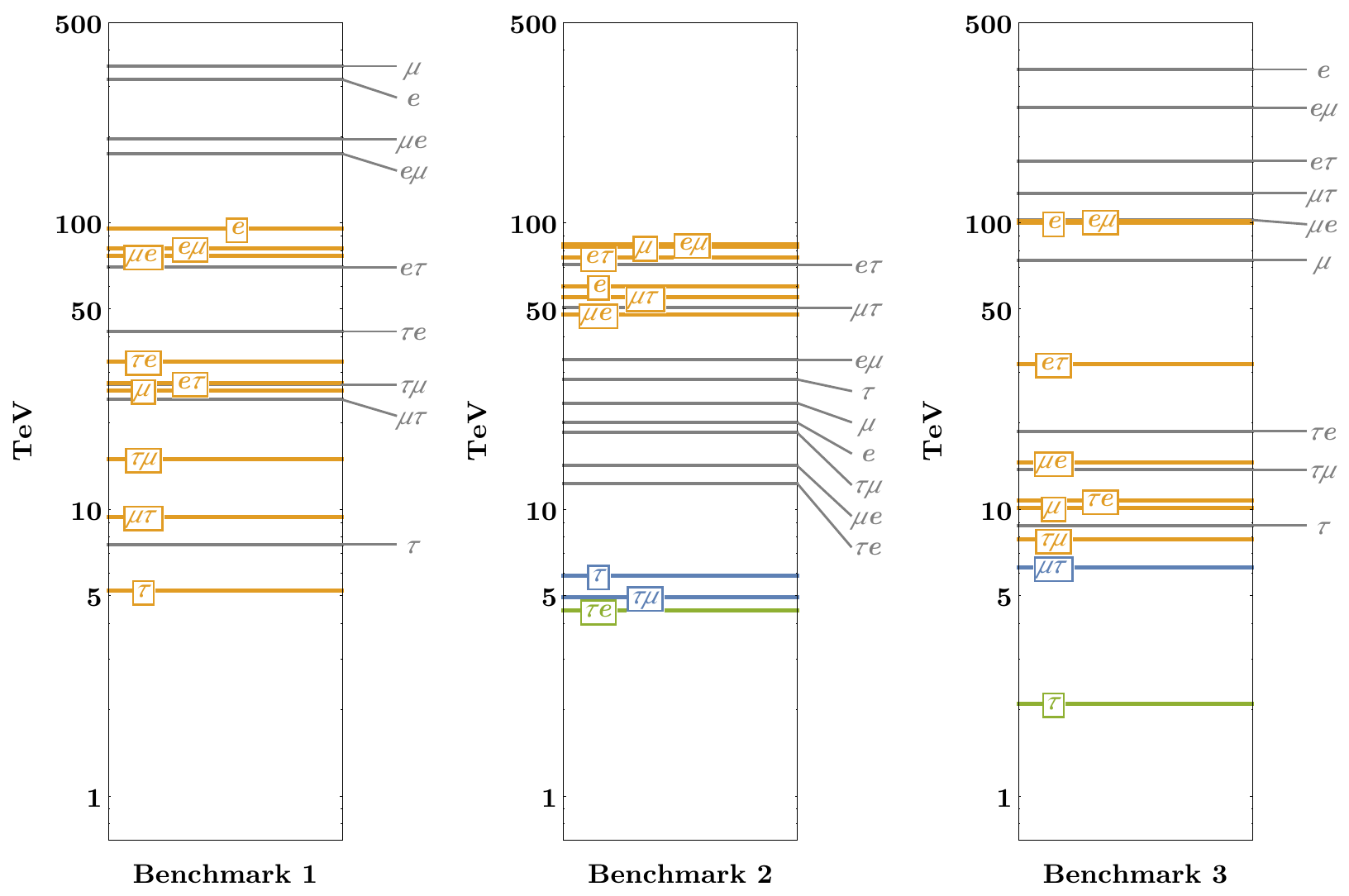}
	\caption{Experimental constraints and mass estimates for the dormant Higgses in the lepton sector in three different benchmarks; the labels denote the indices of the corresponding Higgs. The gray lines are the mass estimates. The colored lines correspond to the most stringent experimental constraint on each of the Higgs masses; orange if the constraint is from $ \mu \to 3e $ decays, blue from $ \tau \to 3\ell $ and green from $ \mu \to e \gamma $.}
	\label{fig:lepton_constraints}\vspace{1cm}
\end{figure*}

Similarly to the quark sector, the only direct constraint on the lepton rotation matrices stem from the requirement $ V_\mathrm{PMNS} = \mathcal{L}_\nu \mathcal{L}_e^\dagger $. 
Therefore, we have the same freedom of choice in their parametrization.
In contrast to the quark sector, the lepton observables do not significantly favor small rotation angles. For our benchmark points we have therefore chosen large rotation matrices to illustrate the case where NP will soon become detectable in LFV observables.\vspace{0.2cm}\\
{\bf Benchmarks}
\begin{alignat}{2}
\text{(1)} \quad \mathcal{L}_{e} &= \mathcal{R}_\nu  = V_\mathrm{PMNS}^{\dagger\; 1/2}, \quad & &  \mathcal{L}_\nu = \mathcal{R}_e = V_\mathrm{PMNS}^{1/2}\no \\ 
\text{(2)} \quad \mathcal{L}_{e} &= \mathcal{R}_\nu  = U_\mathrm{TBM}, \quad & &  \mathcal{L}_\nu = \mathcal{R}_e = V_\mathrm{PMNS} U_\mathrm{TBM} \no\\ 
\text{(3)} \quad \mathcal{L}_{e} &= \mathcal{R}_\nu = V_\mathrm{PMNS}^{\dagger\; 1/2} U_\mathrm{TBM}^{1/2}, \quad & &  \mathcal{L}_\nu = \mathcal{R}_e =V_\mathrm{PMNS}^{1/2}U_\mathrm{TBM}^{1/2}\no
\end{alignat}	
We apply the PDG convention for the PMNS matrix~\cite{Tanabashi:2018oca} using the best-fit values of the three mixing angles and phase, $\delta_{\text{CP}}$~\cite{Esteban:2016qun}. The use of the tribimaximal (TBM) mixing matrix~\cite{Harrison:2002er,Xing:2002sw,Harrison:2002kp,He:2003rm} is motivated from work on discrete flavor model and its structure is given by
\begin{equation}
U_{\text{TBM}} = 
\begin{pmatrix}
-\frac{2}{\sqrt{6}} &  \frac{1}{\sqrt{3}} & 0\\
\frac{1}{\sqrt{6}} & \frac{1}{\sqrt{3}} & \frac{1}{\sqrt{2}} \\
\frac{1}{\sqrt{6}} & \frac{1}{\sqrt{3}} & -\frac{1}{\sqrt{2}} \\
\end{pmatrix}.
\end{equation}

For each of the benchmark points we have determined the strongest individual limit on each $ e $-type dormant Higgs by comparing 
the NP contribution to the decay bounds in \tabref{tab:LFV3b}. We have plotted the leading constraints on the individual Higgs 
masses resulting from our benchmark points in \figref{fig:lepton_constraints} together with the corresponding mass estimates resulting 
from \equaref{eq:mass_estimate}. In the benchmarks almost all the Higgs mass constraints come from the experimental bounds on 
the muon decays---even for the dormant Higgses coupling primarily to the third generation leptons. This is due to the somewhat anarchic 
structure of the PMNS mixing matrix, meaning that all the Higgses mediate LFV muon decays. The PMNS mixing angle suppression  is  insufficient to compensate the weaker bounds on the LFV tau branching ratios. Despite the slightly stronger experimental 
bound on $ \Br(\mu \to e \gamma) $ than $ \Br(\mu^{-} \to e^{-} e^{+} e^{-} ) $, the loop suppression in the leading dormant Higgs 
contribution to the radiative decay is enough that almost all the $ H^e $ Higgses get their strongest bound from the decay to three charged 
leptons. 
For at least some of the Higgses it is possible to engineer the rotation matrices in such a way as 
to avoid the experimental bound on $ \Br(\mu^{-} \to e^{-} e^{+} e^{-} )$.

Regardless of the anarchic flavor structure 
chosen for the benchmark points, \figref{fig:lepton_constraints} shows that there is  only  severe tension between the mass 
estimates of $H_e$, $H_\mu$, $H_{e\mu}$ and $H_{\mu e}$ and their corresponding constraints in Benchmark 2. We therefore expect that most of the parameter space (consistent with 
our assumptions) will pass the lepton-flavor constraints.  
We also anticipate that future CLFV experiments, with significantly improved sensitivities, will be important probes of this scenario.

The Higgses associated with  neutrinos in the Dirac mass scenario, $ H^\nu $, are only constrained by the radiative decays.
In no case, regardless of the left-handed mixing, are the bounds on the $ H^\nu $ masses stronger than $ 5 \, \mathrm{TeV} $, 
much below what is expected from the fine-tuning argument. For this reason the corresponding constraints have not been displayed 
in \figref{fig:lepton_constraints}.

\section{Conclusions}

While our principle of Scalar Democracy lacks a detailed  theoretical underpinning,
it is nonetheless experimentally  testable in the coming years.
If some predictions are confirmed, such as the observation
of $H_b$, and possibly $H_\tau$, $H_{ct}$ and $H_c$, with expected ${\cal{O}}(1)$ Yukawa couplings, it could
reshape our view of the UV completion of the Standard Model.

We propose that a plethora of new scalar bosons exists in nature.
We argue schematically that every fermion pair binds to form a boundstate
scalar boson, due to a universal
attractive interaction at a very high scale, $\Lambda$.  Amongst many new states, including
lepto-quarks, colored isodoublets and singlets, etc., 
this hypothesis implies the existence of a large number of sequentially more massive Higgs bosons.

We argue that the SM Higgs boson is the first of the sequence, and therefore must
have a dynamical origin,  essentially as a $\bar{t}t$ boundstate, but
it is now part of the constellation of composite scalars that influence its mass.
One immediate intrigue is that the
top quark mass, predicted by the renormalization group fixed point \cite{FP1,FP2}, 
which is the prediction of top condensation
models \cite{BHL}, now comes within a few percent of the observed value.
This is expected to be improved by including the multi-Higgs boson masses and decouplings
and is a first indicator that this could be a pathway to new UV physics.

The universal binding we invoke
may be intimately associated with gravity, since
strong scattering amplitudes may exist near the Planck-scale that are
parameterized by $d=6$ operators and scale as
$1/M_{\text{Planck}}^2$.  Amongst these amplitudes we would expected
a general four-fermion structure
such as $(g^2/M^2) (\bar{\psi}_\LL^i \psi_\RR^j) (\bar{\psi}_\RR^i \psi_\LL^j)$,
where $i,j$ run over the conventional quark and lepton flavors and colors
of the SM.
In fact, since gravity does not
distinguish between particle and anti-particle,
we would expect similar operators with a Majorana-like structure. 
The operators must be gauge invariant, and 
with the usual SM fields this contains
a subsector of a global $\SU(48)\times \U(1)$
chiral Lagrangian. 

We therefore argue that there exists a sequence of Higgs doublets leading upwards
to large
mass scales. These  are ``dormant'' isodoublets, each having only
small ``tadpole'' VEVs that arise from mixing with the observed Higgs
and that scale as $1/M^2$ for heavy Higgs fields of mass $M$.
Our main point in invoking
gravity is to motivate that
many Higgs doublets may exist in nature. They could also arise from 
 strong gauge interactions. 
The masses of these scalars are lacking a theory--we view them 
at present as soft-symmetry-breaking, relevant operators
that we insert by hand in accord with known phenomenolgy.

The virtue of the  model  is that 
the Yukawa interactions are ``known,''  and  all
are  of order unity.
This is interesting as it 
transforms the problem of flavor dynamics in a fundamental way: we have only a single
Yukawa coupling at $\Lambda$, and the
derived couplings at low energies are subject to relatively small renormalization corrections. 
The leptons may see the most significant
renormalization to a value of about $0.7$ (smaller `fine-structure'' effects will
be generated by the decoupling of heavy Higgs fields).
The observed spectrum of quarks and leptons and
their mixing angles can be codified in terms of the 
the mass-mixing problem in the 
extended Higgs sector.

Most  interesting is the top-bottom system. The top mass
and the SM Higgs calibrate the overall Yukawa coupling
for all quarks, $g = g_\mathrm{top}\simeq 1$. Here the first sequential dormant
Higgs $H_b$, couples essentially to $\bar{b}b$,  and has a positive mass, $M_b^2$.
It has a mixing with the lightest Higgs 
$H_0$ causes it to acquire a small VEV that feeds 
the $b$-quark its mass.  We can view 
the SM Higgs as having initially  positive $M_0^2$ (or zero), and the  effect
of ``level repulsion'' by $H_b$ drives it to become tachyonic. 
To avoid significant fine tuning we would require $ \mu^4 /M_b^2 \sim 
M_{H}^2 $, which
 leads to a soft ``naturalness'' upper limit on $M_{H_b}\lesssim (v_{\text{weak}}/m_b) (\sim 100$ GeV$)\sim 3.5 
$ TeV.

We discuss in some detail a particular search strategy to address the neutral
component of $H_b=(h_b^+, h_b^0)$, through the process $pp\rightarrow h_b^0\rightarrow \bar{b}b$. This is
available at the current
LHC energy and luminosity, and possible upgrades.  We find that the LHC already has the capability of excluding sequential 
$H_b$ with ${\cal{O}}(1)$
Yukawa coupling to $\bar{b}b$ in a mass range of order $\lta $ 1 TeV (as in the model of \cite{Hill1}).
Our conservative (rough) estimates indicate that the full range of $H_b$ masses
may be accessible to the high luminosity and/or energy-doubled LHC, and should help justify the case for such
future machines.
Our estimates are preliminary and likely will bound above  what can be done
by improved detector based studies and
more detailed deployment of cuts and search strategies. We strongly encourage our experimental
colleagues to pursue this.  We thus defer details of possible searches  $h_b^+\rightarrow t\bar{b}$ and $H_\tau$
elsewhere.

The challenge to Scalar Democracy is
to avoid unwanted $d>4$ flavor transition operators. The rare processes we
focus on are mainly neutral meson mixing, and we find that such processes are binding
on the Higgs spectroscopy. However, avoiding these constraints  in this model is possible
since most of the new sequential Higgs bosons are very heavy.
Conceivably, though we have not explored it, the
extended Higgs sector may provide sources of observable
rare processes in heavy meson decays.
Discovery of the lighter doublets could thus provide
impetus for partially unraveling the flavor problem.

We also discuss the leptons. Our model most naturally leads to a
``neutrino Higgs seesaw'' for neutrino Dirac masses in analogy to the quark-sector 
Higgs structure. However,
we have additional fermionic boundstates that can, in principle, mix to develop tadpole VEVs 
and produce Majorana masses, realizing a type-I or type-II seesaw mechanism.
Hence the double-$\beta$ decay
experiments will be important probes of the far UV Higgsology of this framework.
Our model  can also drive rare lepton number violation experiments,
and we focus discussion on a subset of these, $\ell_i\rightarrow 
\ell_j+\gamma$ and $\ell\rightarrow 3\ell'$.
We certainly do not exclude other possible processes as probes of
the leptonic Higgs system.

One view of the future evolution
of fundamental physics with energy scale, argues that the couplings are small
and asymptotically free,
and that the theory  ``fades away'' into a linear scale-invariance.
We have arrived at  a contrary point of view, that a rich spectrum of new scalar states
lies immediately beyond current energy scales and is within reach of the LHC and its upgrade path.
In part this is motivated by chirality: it is very unlikely that one can
generate a small Yukawa coupling constant from zero (certainly not perturbatively). Hence
the tiny $g_\mathrm{electron} \sim 10^{-6}$ is most likely due to a power law
suppression of a coupling that is of order unity, such as $g_\mathrm{electron} \sim g_\ell \mu_e^2/M^2_e$
where $g_\ell \sim 1$.  The power-law suppression then demands new mass scales, 
such as $\mu$ and $M$, as realized in our model.

We have argued that a natural way to achieve this ``democratically'' throughout the
entire flavor system of the SM is with a grand enlargement of the scalar system.
This flips the flavor problem: the lightest (heaviest) fermions are coupled to the
heaviest (lightest) scalars.  The resulting spectrum of fermions is
due to the inverted spectrum of associated scalars.
This underlies our ``Scalar Democracy.'' 
It focuses urgent attention to the top-bottom (and perhaps the $\tau$--$\nu_\tau$
subsystem) and is testable at the LHC and future upgrades. 
We urge our experimental colleagues and other theorists to take up the cause.

\vspace{0.1 in}

\section*{Acknowledgements}

\noindent
We thank John Campbell, Andre de Gouvea, Roni Harnik, Josh Isaacson and Tobias Neumann for helpful discussions 
and Bogdan Dobrescu and Matthew Low for reading the preliminary draft.
One of us recalls a conversation with Bj, in which he noticed a plethora
of Higgs bosons arising in conjunction with  extensions of gravity,
and which helped to inspire this work \cite{Bjorken}.
This work was done at Fermilab, operated by Fermi Research Alliance,  
LLC under Contract No. DE-AC02-07CH11359 with the United States Department of
Energy.
AET would like to thank Fermilab for hosting him during the writing of this paper, and gratefully acknowledges financial support from the Danish Ministry of Higher Education and Science through an EliteForsk Travel Grant. The CP$^3$-Origins centre is partially funded by the Danish National Research Foundation, grant number DNRF90.

\bibliographystyle{apsrev4-1}
\bibliography{bib}{}
\end{document}